\documentclass{aa}

\usepackage[switch]{lineno}

\usepackage{graphicx}
\graphicspath{ {./Figures/} }
\usepackage{txfonts}
\usepackage{booktabs}
\usepackage{xcolor,chemformula}
\usepackage[separate-uncertainty=true]{siunitx}

\newcommand{\nl}[1]{\textcolor{blue}{(NL: #1)}}
\begin{document}

   \title{The formation of CO$_2$ through consumption of gas-phase CO\\on vacuum-UV irradiated water ice}

   \titlerunning{CO$_2$ production through gas-phase CO consumption on VUV irradiated water ice}

   \subtitle{}

   \author{J. Terwisscha van Scheltinga\inst{1,2,\thanks{Current address: Department of Astronomy, University of Virginia, P.O. Box 400325, Charlottesville, VA 22904, USA}}
          \and
          N.F.W. Ligterink\inst{3}\
          \and
          A.D. Bosman\inst{4}
          \and
          M. R. Hogerheijde\inst{2,5}
          \and
          H. Linnartz\inst{1}
          }
   \institute{Laboratory for Astrophysics, Leiden Observatory, Leiden University, P.O. Box 9513, 2300 RA Leiden, The Netherlands\\
              \email{jeroentvs@virginia.edu}
            \and
            Leiden Observatory, Leiden University, P.O. Box 9513, 2300 RA Leiden, The Netherlands
            \and
            Physics Institute, University of Bern, Sidlerstrasse 5, 3012 Bern, Switzerland
            \and
            Department of Astronomy, University of Michigan, 323 West Hall, 1085 S. University Avenue, Ann Arbor, MI 48109, USA
            \and
            Anton Pannekoek Institute for Astronomy, University of Amsterdam, Science Park 904, 1098 XH Amsterdam, The Netherlands 
             }

   \date{Received 8 September 2021 / Accepted 16 August 2022}

 
  \abstract
   {Recent observations of protoplanetary disks suggest that they are depleted in gas-phase CO up to a factor of 100 with respect to predictions from physical-chemical (or thermo-chemical) models. It has been posed that gas-phase CO is chemically consumed and converted into less volatile species through gas-grain processes. Observations of interstellar ices reveal a CO$_2$ component in a polar (H$_2$O) ice matrix, suggesting potential co-formation or co-evolution.}
   {The aim of this work is to experimentally verify the interaction of gas-phase CO with solid-state OH radicals on the surface of water ice above the sublimation temperature of CO.}
   {Amorphous solid water (ASW) is deposited in an ultra-high vacuum (UHV) setup at 15~K and irradiated with vacuum-UV (VUV) photons (140--170~nm, produced with a microwave-discharge hydrogen-flow lamp) to dissociate H$_2$O and create OH radicals. Gas-phase CO is simultaneously admitted and only adsorbs with a short residence time on the ASW. Formed products in the solid state are studied in the infrared through Fourier transform infrared spectroscopy and once released into the gas phase with quadrupole mass spectrometry.}
   {Our experiments show that gas-phase CO is converted into CO$_2$ when interacting with ASW that is VUV irradiated with a conversion efficiency of 7--27\%. Between 40 and 90~K, CO$_2$ production is constant, above 90~K, CO$_2$ production is reduced in favor of O$_2$ production. In the temperature range of 40--60~K, the CO$_2$ remains in the solid state, while at temperatures $\geq$ 70~K the majority of the formed CO$_2$ is immediately released into the gas phase.}
   {We conclude that gas-phase CO reacts with OH radicals, created on the surface of ASW with VUV irradiation, above its canonical sublimation temperature. The diffusion during the short, but nonzero, residence times of CO on the surface of ASW suggests that a Langmuir-Hinshelwood type reaction is involved.
   This gas-phase CO and solid-state OH radical interaction could explain (part of) the observed presence of CO$_2$ embedded in water-rich ices when it occurs during the build up of the H$_2$O ice mantle. It may also contribute to the observed lack of gas-phase CO in planet-forming disks, as previously suggested. It should be noted though that our experiments indicate a lower water ice dissociation efficiency than originally adopted in model descriptions of planet-forming disks and molecular clouds. Incorporation of the reduced water ice dissociation and increased binding energy of CO on a water ice surfaces in physical-chemical models would allow investigation of this gas-grain interaction to its full extend.}


   \maketitle
%
\section{Introduction}
In typical laboratory astrochemistry experiments, the processes that occur in the solid state and gas phase are investigated independently. However, there are conditions in the interstellar medium where these are intimately intertwined and could affect each other. In this work, we explore experimentally the interaction between gas-phase carbon monoxide (CO) and UV irradiated water (H$_2$O) ice and place the results in astrophysical context.

In the study of planet forming disks, CO and its isotopologues are common tracers of the total gas mass, but are often found to be depleted by factors up to 10--100, even after taking into account freeze-out of CO in the coldest disk regions \citep{2016_Ansdell_Lupus_ApJ...828...46A, 2017_Miotello_Lupus_A&A...599A.113M, 2021_Trapman_CO_conv_A&A...649A..95T}. Recent physical-chemical models suggest that gas-phase CO could be converted into CO$_2$ after interaction with a UV-irradiated H$_2$O ice surface, at temperatures just above the CO sublimation temperature \citep{2016_Drozdovskaya_midplanes_MNRAS.462..977D, 2016_Eistrup_ChemEvo_A&A...595A..83E, 2018_Bosman_CO_A&A...618A.182B}. Under realistic disk conditions, this pathway was found by \citet{2018_Bosman_CO_A&A...618A.182B} to convert significant amounts of gas-phase CO into CO$_2$. However, little experimental work exists to confirm this process. If efficient, the UV-irradiated edges of molecular clouds could be another environment where this gas-grain reaction can occur. This possibly explains (part of) the observed presence of CO$_2$ in polar ices \citep[see e.g.,][]{1999_Gerakines_ISO_CO2_ApJ...522..357G, 2008_Pontoppidan_CO2_ApJ...678.1005P}, if gas-phase CO conversion happens on the grain surface during the build up of the H$_2$O ice mantle, adding to contributions from other CO$_2$ formation pathways already studied.

The solid-state formation of CO$_2$ has been investigated both theoretically \citep[see e.g.,][]{2008_Goumans_CO2_comp_MNRAS.384.1158G, 2010_Goumans_CO_O_MNRAS.406.2213G, 2013_Arasa_HOCO_JPCA..117.7064A} and experimentally. Several energetic and non-energetic pathways have been experimentally confirmed to form CO$_2$ under astrophysical conditions, such as, ground-state CO reacting with an electronically excited CO* to form CO$_2$ and atomic carbon \citep{1996_Gerakines_UV_ices_A&A...312..289G, 1998_Palumbo_CO2_A&A...334..247P, 2005_Loeffler_CO2_A&A...435..587L, 2006_Jamieson_CO_10K_ApJS..163..184J, 2009_Bennett_CO2_PCCP...11.4210B, 2009_Ioppolo_CO2_A&A...493.1017I}, CO reacting with atomic oxygen to form CO$_2$ \citep{2001_Roser_CO+O_ApJ...555L..61R, 2006_Madzunkov_CO+O_PhRvA..73b0901M, 2011_Raut_CO_O_ApJ...737L..14R, 2013_Ioppolo_RScI...84g3112I, 2013_Minissale_CO_O_A&A...559A..49M}, CO reacting with a hydroxyl (OH) radical to form CO$_2$ and atomic hydrogen \citep{2002_Watanabe_CO2_ApJ...567..651W, 2007_Watanabe_H2O-CO_ApJ...668.1001W, 2009_Ioppolo_CO2_A&A...493.1017I, 2010_Oba_non_OH_CO_ApJ...712L.174O, 2011_Oba_CO_OH_40-60K_PCCP...1315792O, 2011_Ioppolo_CO2_MNRAS.413.2281I, 2011_Noble_CO2_ApJ...735..121N, 2011_Zins_CO+OH_ApJ...738..175Z, 2014_Yuan_ERCO2_ApJ...791L..21Y}, formaldehyde (H$_2$CO) reacting with atomic oxygen and to form CO$_2$ and molecular hydrogen \citep{2015_Minissale_CO2_A&A...577A...2M}, or oxidation of carbonaceous surfaces \citep{2006_Mennella_CO2_carbon_ApJ...643..923M, 2012_Fulvio_CO2_carbon_ApJ...752L..33F, 2012_Raut_CO2_from_ASW_carbongrains_ApJ...752..159R, 2015_Sabri_CO2_carbon_A&A...575A..76S, 2015_Shi_Oxi_Graphite_ApJ...804...24S}.

The above pathways were found to significantly produce CO$_2$ in the solid state and only the last two pathways do not include CO. The majority of these solid-state experiments are performed at temperatures below 20~K, representative of dark cloud or disk midplane ($>20$~AU) conditions. This is well below the CO sublimation temperature, which is approximately 20 and 30~K for interstellar and laboratory timescales, respectively \citep[see e.g.,][]{2011_Fayolle_CO_PSD_ApJ...739L..36F, 2016_Schwarz_CO_ApJ...823...91S}. The low temperatures in these experiments ensure that CO stays adsorbed on the surface and is able to react with the other ice constituents. The experimental studies by \citet{2011_Oba_CO_OH_40-60K_PCCP...1315792O} and \citet{2014_Yuan_ERCO2_ApJ...791L..21Y} have investigated the formation of CO$_2$ from CO at substrate temperatures above the sublimation temperature of CO. In the former, CO$_2$ was formed when CO and OH radicals were co-deposited on a substrate in the temperature range from 40 to 60~K. The latter observed formation of CO$_2$ when gas-phase CO interacted with OH radicals produced by UV photons on the surface of water ice at 76~K. Both works show that CO can interact with OH radicals in the solid state above its canonical desorption temperature.

In this work, we set out to experimentally investigate the conversion of gas-phase CO into CO$_2$ on the surface of vacuum-UV (VUV) irradiated water ice (40--120~K), and assess the efficiency in astrophysical settings. Specifically, amorphous solid water (ASW) is irradiated at a temperature of $\geq40$~K, which ensures that the majority of the gas-phase CO in our experimental chamber does not freeze out onto our ASW sample. Section~\ref{sec:methods} describes the methods used to investigate this process, and analyze the data. Results are presented in Sect.~\ref{sec:results} and are discussed in Sect.~\ref{sec:disc}. The astrophysical implications are given in Sect.~\ref{sec:astro_imp}, and concluding remarks are summarized in Sect.~\ref{sec:conc}.


\begin{table*}
    \caption{Overview of performed experiments} \label{tab:exp}
    \centering
    \begin{tabular}{llccc}
        \toprule\toprule
        Series  & \multicolumn{1}{c}{Molecules} & Temperature$^a$   & \ch{H2O} column density$^{b, c}$   & Notes\\
                &                               & (K)           & (monolayers)              & \\
        \midrule
        Main experiments    & H$_2${}$^{18}$O (s) + $^{13}$C$^{18}$O (g)  &  \phantom{1}40\phantom{1}   & 57.8 & -- \\
                            & H$_2${}$^{18}$O (s) + $^{13}$C$^{18}$O (g)  &  \phantom{1}50\phantom{1}   & 56.7 & -- \\
                            & H$_2${}$^{18}$O (s) + $^{13}$C$^{18}$O (g)  &  \phantom{1}60\phantom{1}   & 57.6 & -- \\
                            & H$_2${}$^{18}$O (s) + $^{13}$C$^{18}$O (g)  &  \phantom{1}70\phantom{1}   & 61.8 & -- \\
                            & H$_2${}$^{18}$O (s) + $^{13}$C$^{18}$O (g)  &  \phantom{1}80\phantom{1}   & 62.9 & -- \\
                            & H$_2${}$^{18}$O (s) + $^{13}$C$^{18}$O (g)  &  \phantom{1}90\phantom{1}   & 57.8 & -- \\
                            & H$_2${}$^{18}$O (s) + $^{13}$C$^{18}$O (g)  &  100\phantom{1}             & 56.8 & -- \\
                            & H$_2${}$^{18}$O (s) + $^{13}$C$^{18}$O (g)  &  120\phantom{1}             & 58.3 & -- \\
        \midrule
        Control             & H$_2${}$^{18}$O (s)                         & \phantom{1}40\phantom{1}    & 70.2 & \multicolumn{1}{l}{water only} \\
                            & H$_2${}$^{18}$O (s) + $^{13}$CO (g)         & \phantom{1}60\phantom{1}    & 63.2 & \multicolumn{1}{l}{{$^{13}$C$^{16}$O}} \\
                            & H$_2${}$^{18}$O (s) + $^{13}$C$^{18}$O (g)  & \phantom{1}40\phantom{1}    & 66.4 & \multicolumn{1}{l}{no VUV irradiation} \\
        \bottomrule
    \end{tabular}
    \tablefoot{$^{(a)}$ All ices have been deposited at 15~K. The temperature refers to the value at which the ASW is VUV irradiated. $^{(b)}$ The H$_2$O column density is derived through the integrated area of the OH-stretching mode (boundaries, 3800--2950~cm$^{-1}$) through Eq.~\ref{eq:col_den}, where $A'$ is taken to be $1.5 \times 10^{-16}$~cm molec$^{-1}$ \citep[H$_2${}$^{16}$O,][]{2015_Bouilloud_A'_MNRAS.451.2145B}. $^{(c)}$ It is likely that this value represents a lower limit due to the nonlinearity of RAIRS at column densities above 10~monolayers.}
\end{table*}

\section{Methods} \label{sec:methods}
\subsection{CryoPAD2}
All reported laboratory measurements are performed in the Leiden Laboratory for Astrophysics using the Cryogenic Photoproduct Analysis Device 2 \citep[CryoPAD2;][]{2017_Niels_CH3NCO_MNRAS.469.2219L,2018_Niels_CH3OH_A&A...612A..88L}. This setup operates under ultra-high vacuum conditions ($\mathrm{P_{mc}} \sim 5 \times 10^{-11}$~mbar at 15~K). It accommodates a gold-coated substrate which is positioned in the center of a stainless steel chamber and acts as an analogue for an interstellar dust-grain surface. On top of the chamber a closed-cycle helium cryostat is positioned which cools the gold-coated surface down to temperatures of 15~K. A Lakeshore 350 temperature controller sets the temperature of the substrate through PID-controlled Joule heating in the range of 15 to 300~K with an absolute and relative accuracy of $\pm2$ and $\pm1$~K, respectively. In order to further simulate the interstellar environments in which these dust grains reside, a microwave-discharge hydrogen-flow lamp (MDHL) is connected to the chamber. These type of sources generally produce VUV photons at 121.6~nm, Lyman-$\alpha$, and between 140 to 170~nm, which corresponds to photon energies of 7.5 to 10.2~eV. However, in the present experiments a MgF$_2$ window is used which absorbs Lyman-$\alpha$ photons, see Appendix \ref{app:uv_spec} for the VUV spectrum. The flux of the MDHL at the location of the substrate is determined with a NIST-calibrated photodiode (SXUV-100) as $(2.5\pm0.3) \times 10^{14}$~photons s$^{-1}$ cm$^{-2}$.

The reactions induced by VUV irradiation under these conditions are diagnosed using infrared spectroscopy and mass spectrometry. The collimated beam of a Fourier-Transform InfraRed Spectrometer (Agilent 660 FTIRS), is used for Reflection Absorption InfraRed Spectroscopy (RAIRS). In this method the incoming FTIR beam is reflected from the substrate under a grazing incidence angle, improving the sensitivity. This in situ diagnostic allows us to probe, qualitatively and quantitatively, the molecular content in the ice adsorbed on the substrate. The infrared spectra are acquired continuously during the experiments to investigate and track the chemical evolution in the solid state under the influence of VUV irradiation. 

The second diagnostic tool is a Hiden HAL/3F PIC 1000 series quadrupole mass spectrometer (QMS). During VUV irradiation some molecular species desorb from the substrate into the gas phase. The QMS probes the molecular content of the atmosphere in the chamber. This allows for qualitative assignment of species released or produced during the experiments through their characteristic mass-fragmentation patterns. Furthermore, after calibration of the QMS through the procedure described in Sect.~\ref{ssec:QMS_cal}, it is possible to derive the quantitative amount of a species released into the gas phase. After VUV irradiation, the substrate temperature is linearly increased with time in a temperature programmed desorption (TPD) experiment until all adsorbed species have thermally desorbed. During TPD, species are released into the gas phase at their canonical desorption temperature, and are subsequently measured by the QMS. Upon ionization, in our case with 70~eV electrons, molecules fragment into a characteristic fragmentation pattern, which allows for assignment of newly formed species, complementary to the infrared. Rare isotopologue precursors are used to discriminate from background gas contaminations, and to add diagnostic information to the RAIRS and TPD experiments.

\subsection{Experimental protocol}
The following molecules are used in the experiments: Milli-Q H$_2$O (Type I), H$_2${}$^{18}$O (Sigma-Aldrich, 97\%), regular CO (Linde gas, 99.997\%), and $^{13}$C$^{18}$O (Sigma-Aldrich, 99\% $^{13}$C and 95\% $^{18}$O). The experiment is started by depositing a layer of ASW onto the substrate. The gas-phase H$_2$O enters the chamber roughly one~cm away from the substrate, and is deposited under normal incidence to the substrate through a capillary array. The temperature of the substrate during H$_2$O deposition is set at 15~K and the water ice is deposited for 10 minutes. This ensures that the deposited H$_2$O is porous-ASW. A precision leak valve is used to guarantee consistent column densities of H$_2$O throughout the experiments. Before the experiments continue, the chamber is left to settle for at least 30~minutes. This ensures that the pressure in the main chamber (P$_{\mathrm{mc}}$) is below $2.0 \times 10^{-10}$~mbar and that the amount of residual gas-phase H$_2$O can be neglected. After this, the substrate temperature is slowly increased, with a rate of 2~K min$^{-1}$, to the temperature at which the experiments are performed. Once the desired sample temperature is reached, the MDHL is started and gas-phase CO is admitted into the chamber. The gas-phase CO enters the chamber at roughly 5 cm distance from the substrate, and under 45~degrees with respect to the substrate normal. The precision leak valve is set to have a constant P$_{\mathrm{mc}}$ of $5.0 \times 10^{-8}$~mbar. This translates into the ASW surface being exposed to approximately $5 \times 10^{13}$~CO molecules cm$^{-2}$ s$^{-1}$. The ASW is exposed in total 300~minutes to VUV irradiation (with a total incident fluence of $4.5 \times 10^{18}$~photons) and CO molecules, after which TPD is performed to sublimate parent and newly-formed species. During VUV irradiation, the shutter between the MDHL and vacuum chamber is closed periodically to measure the baseline signals of the chamber without VUV irradiation. The experiments performed in this study are listed in Table~\ref{tab:exp}.

\subsection{Data analysis} \label{sec:analysis}
\subsubsection{RAIRS vibrational spectroscopy} \label{ssec:RAIRS}
The infrared spectra are acquired in RAIRS mode with the FTIR and are subsequently analysed. The column density, $N_{\rm{species}}$, of the probed molecules on the substrate is derived through the following relationship with the measured absorbance
\begin{equation} \label{eq:col_den}
    N_{\rm{species}} = \ln(10)\cdot\frac{\int_{band}\log_{10}\left(\frac{I_0(\tilde{\nu})}{I(\tilde{\nu})}\right) d\tilde{\nu}}{R\cdot A'},
\end{equation}
where the absorbance, the ratio of the incoming flux, $I_0(\tilde{\nu})$, and reflected flux, $I(\tilde{\nu})$, is integrated over a range that encompasses the full absorption feature, and $A'$ is the apparent band strength. The apparent band strengths are taken from literature from transmission experiments; for RAIRS these values need to be corrected with a value $R$ in order to retrieve accurate column densities. The RAIRS correction factor (\textit{R}) is empirically determined on CryoPAD2 through isothermal desorption of CO and is found to be 4.5 \citep[see e.g.,][for a description of the used approach]{2009_Oberg_CO_PSD_A&A...496..281O, 2018_Ligterink_peptide_MNRAS.480.3628L}. We assume that the area probed by the infrared beam on our substrate amounts to 1.0~cm$^2$ and thus the amount of molecules, $N_{\rm{species}}$, is also the column density, given in molecules cm$^{-2}$. As stated before and shown in Table~\ref{tab:exp}, all main experiments are performed with $^{13}$C$^{18}$O and H$_2$$^{18}$O resulting in the formation of $^{13}$C$^{18}$O$_2$. To our knowledge the apparent band strength of this specific isotope of CO$_2$ is unknown and thus the apparent band strength of $^{13}$CO$_2$, $6.8 \times 10^{-17}$~cm molecule$^{-1}$ \citep{2015_Bouilloud_A'_MNRAS.451.2145B}, is used to approximate the column density of $^{13}$C$^{18}$O$_2$.

As is shown in Sect.~\ref{sec:results}, multiple CO$_2$ features are observed in the infrared. In order to follow the growth of these different CO$_2$ features, the three prominent ones are approximated by fitting a Gaussian profile to each of them in order to deconvolve the spectra. The \textsc{curve\_fit} function from \textsc{SciPy} is used to fit a Gaussian profile to each absorption component through least squares regression \citep{2020SciPy-NMeth}. A 3-Gaussian fit reproduces the integrated absorbance to $\leq10\%$, and suffices as a fit, given the variation in observed profile shapes due to (small) changes in physical or chemical environment and spectroscopic artifacts in the spectra.

\subsubsection{QMS calibration} \label{ssec:QMS_cal}
RAIRS allows determination of the column density in the solid state, while the QMS allows for quantification of molecules released from the solid state into the gas phase. In order to use the QMS for quantitative purposes, mass signals need to be calibrated and this is realized, through the photodesorption of CO \citep[see e.g.,][]{2013_Fayolle_PD_N2_O2_A&A...556A.122F, 2015_Martin_CO2_A&A...584A..14M}. The loss of solid-state CO is traced with RAIRS and is correlated to the gas-phase CO signal measured simultaneous by the QMS. This calibration allows for the conversion of any measured gas-phase QMS signal, released under VUV irradiation, to a column density from the solid state. However, one needs to correct for the difference in the electron-impact ionization cross section, the fractional fragmentation, and the mass sensitivity of the QMS of the investigated molecule with respect to CO. The following equation is used to quantify the amount of CO$_2$ formed in the solid state and subsequently released into the gas phase
\begin{linenomath}
\begin{equation} \label{eq:QMS_cal}
    \frac{N_{\rm{CO_2(ice)}}}{\int S_{\rm{CO_2(gas)}}} = \frac{\sigma_{\rm{CO}}}{\sigma_{\rm{CO_2}}} \cdot \frac{F(\rm{CO^+/CO})}{F(\rm{CO{_2}^{+}/CO_2})} \cdot \frac{M(\rm{CO})}{M(\rm{CO_2})} \cdot \frac{N_{\rm{CO(ice)}}}{\int S_{\rm{CO(gas)}}},
\end{equation}
\end{linenomath}
where $N_{\rm{CO_2}}$ is the column density of CO$_2$ released from the solid state, $\int S_{\rm{CO_2}}$ the integrated CO$_2$ QMS signal, $\sigma$ the total electron-impact ionization cross section, $F$ the fragmentation fraction of the ionized species, and $M$ the mass sensitivity function of the Hiden QMS on CryoPAD2. The last term of Eq.~\ref{eq:QMS_cal} is the CO calibration experiment, where $N_{\rm{CO}}$ is the amount of CO that photodesorbed from the solid state and $\int S_{\rm{CO}}$ the integrated CO signal measured by the QMS during photodesorption. In a similar fashion the column density of O$_2$ is determined, but with its respective parameters. The electron-impact ionization cross sections ($\sigma$) used in this work for CO, CO$_2$, and O$_2$ are 2.516, 3.521, and 2.441~$\AA^2$ at 70~eV, respectively \citep{EIICS_database}. The fragmentation fractions (\textit{F}) for CO$^+$, CO$_2$$^+$, and O$_2$$^+$ are 0.940, 0.942, and 0.821 respectively, where unity is the normalized summation of all fragmentation fragments.

\begin{figure}[t!]
    \includegraphics[width=\hsize]{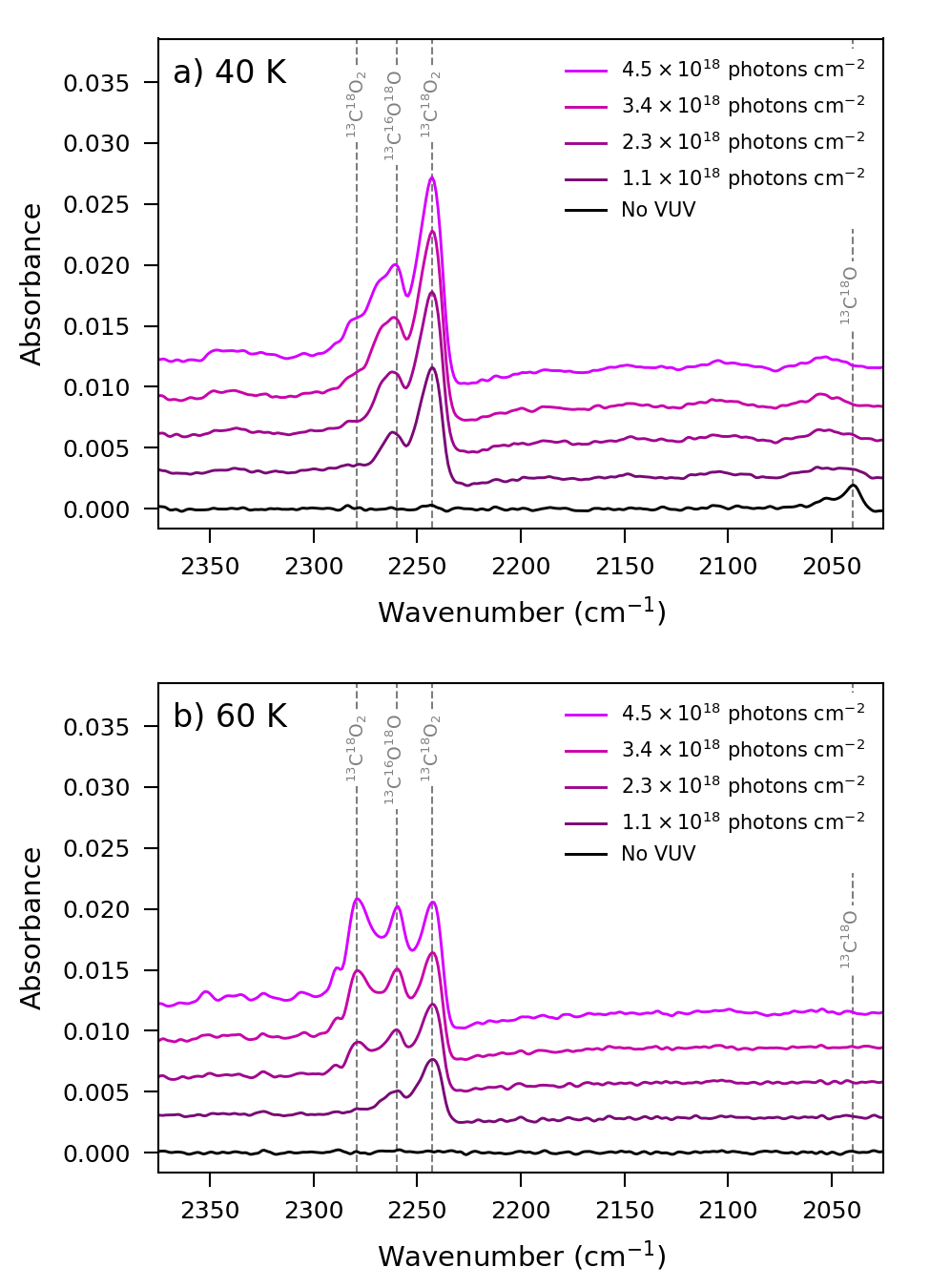}
    \caption{Five difference RAIRS spectra following VUV irradiation of ASW (H$_2$$^{18}$O) at 40~K (top) and 60~K (bottom) with gas-phase $^{13}$C$^{18}$O at different VUV fluences, increasing from low to high. The vertically dashed lines indicate the assignment of the main absorption features.} \label{fig:diff_spectra}
\end{figure}

\section{Results} \label{sec:results}
In this section, we present the results of the experiments mentioned in Table~\ref{tab:exp}. In general, irradiation of ASW with VUV photons in the presence of gas-phase CO produces CO$_2$. Additionally, in the experiments at the higher end of the temperature range, ($>$ 90~K), formation of molecular oxygen (O$_2$) is observed. To understand the processes that occur on or in the solid state we consider the infrared and QMS results, and how these change with temperature.

\begin{figure*}[t!]
    \includegraphics[width=\hsize]{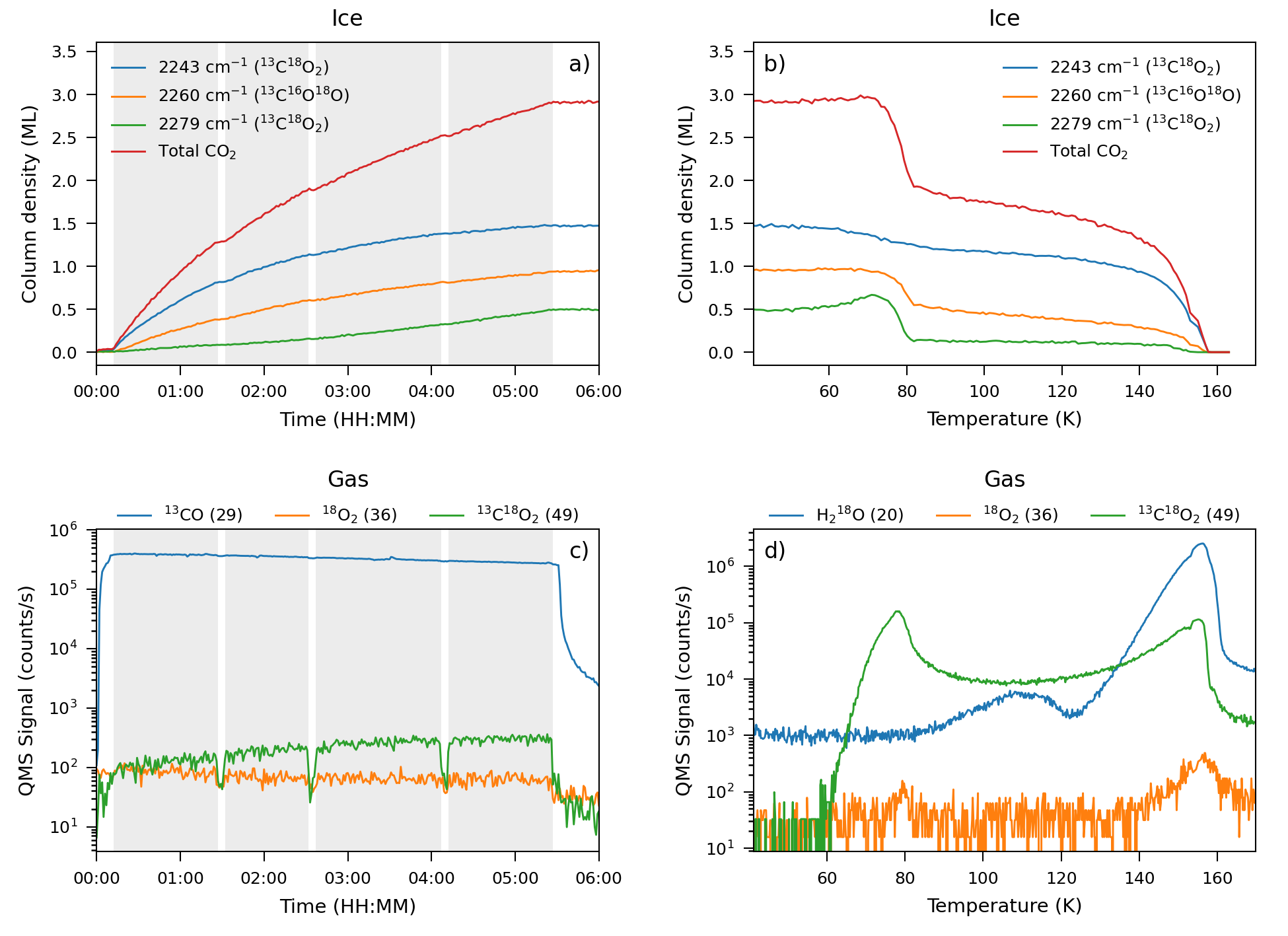}
    \caption{Results of the experiment of ASW at a temperature of 40 K. In the top row the deconvolved infrared components, which trace the solid state, are given, based on RAIRS during VUV irradiation (panel a) and temperature programmed desorption (TPD) in panel b). In the bottom row the data from the quadrupole mass spectrometer (QMS) is presented, which trace the gas phase. The left column shows data acquired during VUV irradiation (shaded areas indicate when the VUV shutter is open) and on the right during TPD. In panel c) $^{13}$CO is shown, as measuring the main isotopologue ($^{13}$C$^{18}$O) would saturate the QMS. The $^{16}$O isotope is present in the CO sample at a level of 5\%.}
    \label{fig:exp_40K}
\end{figure*}

\subsection{Infrared spectroscopy results} \label{ssec:infrared}
We observe the formation of CO$_2$ in the solid state through RAIRS. In Fig.~\ref{fig:diff_spectra} we present five difference RAIRS spectra at experimental temperatures of 40 and 60~K, in the top and bottom panel, respectively. Such spectra are obtained by subtracting the initial ASW spectrum, before VUV irradiation, from the subsequently acquired spectra during irradiation. The spectra shown here are obtained at five different VUV fluence intervals where ASW was simultaneously exposed to gas-phase CO. Although the main isotopes used are $^{13}$C and $^{18}$O, small amounts of $^{12}$C and $^{16}$O are present in our samples. It is apparent that in the wavenumber range 2350--2225~cm$^{-1}$ absorption features grow with increasing fluence. All of the features in this range are attributed to isotopologues of CO$_2$. The three most distinct features are positioned at 2279, 2260, and 2243~cm$^{-1}$ and are attributed to $^{13}$C$^{18}$O$_2$ aggregates on top of the water ice, $^{13}$C$^{16}$O$^{18}$O bound to water, and $^{13}$C$^{18}$O$_2$ bound to water, respectively \citep{1977_Lehmann_CO2_ApPhy..13..153L,2017_Jiao_CO2_ApJ...837...65H}.

The lowest ASW temperature at which the experiments are performed is 40~K. This ensures that the majority of the gas-phase CO molecules that enter the vacuum chamber cannot adsorb onto our sample, as it is above the canonical desorption temperature of CO. However, as shown by \citet{2016_Jiao_stick_ApJ...823...56H}, the sticking coefficient of CO on nonporous-ASW (np-ASW) is close to unity at 40~K. Once the ASW is covered with CO, no additional CO freeze-out occurs. This is also seen in our experiment at 40~K through the infrared signal around 2040~cm$^{-1}$ where, preceding VUV-irradiation, the ASW is briefly exposed to gas-phase CO only (Fig.~\ref{fig:diff_spectra}a). In this short 5~minute window, CO adsorbs on top of the ASW with a column density of $\sim1.1$~monolayers, where one monolayer equals $10^{15}$~molecules cm$^{-2}$. This CO ice grows within 60~s and does not further increase. As soon as VUV irradiation starts, this solid-state CO on the surface of the ASW is converted into CO$_2$. At ASW temperatures $\geq50$~K no adsorption of CO is seen, illustrated here for a temperature of 60~K (Fig.~\ref{fig:diff_spectra}b). This is expected as the sticking coefficient of CO on ASW significantly drops at temperatures $\geq50$~K. An upper limit of $\leq0.1$~monolayers is derived for CO adsorbed on top of ASW at temperatures $\geq50$~K.

Figure~\ref{fig:exp_40K} shows the combined results of the experiment with ASW at 40~K. The upper panels show infrared data  during irradiation (a) and after irradiation upon TPD (b). The lower panels (c and d) show the corresponding gas-phase data recorded with the QMS. Figure~\ref{fig:exp_40K}a shows the growth of each individual solid-state CO$_2$ component during VUV irradiation (shaded areas) as well as the combined results (red curve), while Fig.~\ref{fig:exp_40K}b shows the subsequent decrease during TPD after irradiation. It is evident that the deconvolved components evolve differently from each other. The CO$_2$ component at 2243~cm$^{-1}$ is the first to grow and levels off as the VUV fluence increases. This component is attributed to $^{13}$C$^{18}$O$_2$ that initially forms and is bound to the ASW surface. Sequentially, aggregates start to form on top of the water ice as the column density of CO$_2$ increases, because CO$_2$ does not wet the ASW surface, that is, the binding energy between CO$_2$--CO$_2$ is higher than CO$_2$--H$_2$O \citep{2017_Jiao_CO2_ApJ...837...65H}. These CO$_2$ aggregates absorb infrared light at a different wavenumber, namely 2279~cm$^{-1}$ \citep[cf.][who observed the same features, but shifted by $\sim100$ cm$^{-1}$ due to the different isotopologue used]{2017_Jiao_CO2_ApJ...837...65H}. With increasing temperature of the ASW, the diffusion of CO$_2$ across the ASW increases. This increased diffusion results in earlier formation of CO$_2$ aggregates and the amount of molecules in these aggregates increases. This is confirmed by less CO$_2$ molecules bound to ASW in the 2243~cm$^{-1}$ component with increasing temperature, see Figs.~\ref{fig:exp_40K}a (40~K), \ref{fig:exp_50K}a (50~K), and \ref{fig:exp_60K}a (60~K). The absorption feature at 2260~cm$^{-1}$ is due to $^{13}$C$^{16}$O$^{18}$O, formed from isotope impurities, bound to the ASW surface and is super imposed on top of the two absorption features of $^{13}$C$^{18}$O$_2$. There is a nonzero baseline between the 2243 and 2279~cm$^{-1}$ features due to the range of binding energies on the surface of H$_2$O with CO$_2$. As a result, the fitting of the 2260~cm$^{-1}$ component comprises contributions from both the 2243 and 2279~cm$^{-1}$ features. This makes an unique assignment and quantification of the 2260~cm$^{-1}$ component difficult.

A decrease in the CO$_2$ column density is shown in Fig.~\ref{fig:exp_40K}b, following sublimation into the gas phase during TPD. The overall decrease as well as the decrease of the individual components are shown. The CO$_2$ sublimates in two steps, the first desorption event occurs at $\sim80$~K and the second at $\sim155$~K. The former is in line with the canonical desorption temperature of CO$_2$, and the latter with the canonical desorption temperature of H$_2$O. The component at 2279 cm$^{-1}$ drops around $\sim80$~K, which is in line with aggregates of CO$_2$ on top of the water ice. The component at 2243~cm$^{-1}$ gradually drops as the temperature of the ASW increases and disappears with the desorption of H$_2$O at 156~K, which is in agreement with CO$_2$ bound to ASW surface.

\begin{figure*}[t!]
    \includegraphics[width=\hsize]{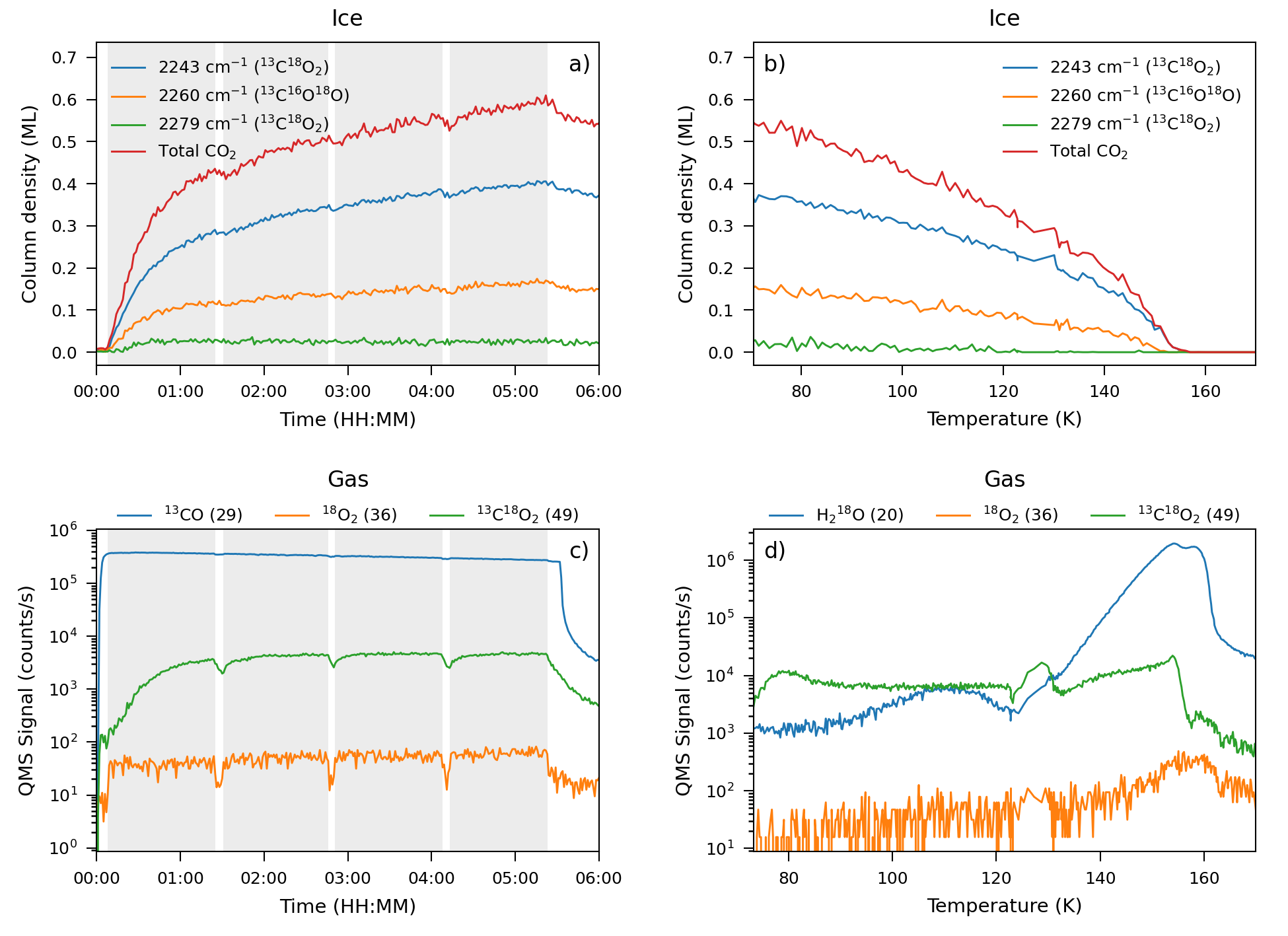}
    \caption{Results of the experiment of ASW at a temperature of 70~K. In the top row the deconvolved infrared components, which trace the solid state, are given, based on RAIRS during VUV irradiation (panel a) and temperature programmed desorption (TPD) in panel b). In the bottom row the data from the quadrupole mass spectrometer (QMS) is presented, which trace the gas phase. The left column shows data acquired during VUV irradiation (shaded areas indicate when the VUV shutter is open) and on the right during TPD. In panel c) $^{13}$CO is shown, as measuring the main isotopologue ($^{13}$C$^{18}$O) would saturate the QMS. The $^{16}$O isotope is present in the CO sample at a level of 5\%. The TPD QMS data in panel d) between 120--130~K is unreliable due to a nonlinear temperature artifact during heating of the sample.}
    \label{fig:exp_70K}
\end{figure*}

At ASW temperatures $\geq70$~K, the majority of the formed CO$_2$ is immediately released back into the gas phase (see e.g., Fig.~\ref{fig:exp_70K}). At 70 and 80~K, however, some of the initially formed CO$_2$ remains in the solid state, see Fig.~\ref{fig:exp_70K}a and Appendix~\ref{fig:exp_80K}a, respectively. This CO$_2$ is formed during the initial moments of VUV irradiation and is bound to the deep binding sites on the ASW surface that are able to ``trap'' CO$_2$. The column densities of this CO$_2$ at 70 and 80~K are 0.6 and 0.12~monolayers, respectively, compared to the $\sim3$~monolayers formed at temperatures below 70~K. No solid-state CO$_2$ is detected in the experiments with ASW temperatures $\geq90$~K, and the upper limit of solid-state CO$_2$ is derived to be $\leq0.02$~monolayers.

\subsection{QMS results} \label{ssec:QMS}
The QMS allows for gas-phase species to be traced in the chamber during VUV irradiation and afterwards during TPD. The signals measured during TPD are only used for identification. In the following two sections we focus first on the QMS analysis of CO$_2$ that remained in the solid state (40--60~K) and then on CO$_2$ released into the gas phase ($\geq70$~K) after formation.

\subsubsection{Solid-state CO$_2$ (40--60~K)}
The majority of the CO$_2$ formed at ASW temperatures of 40--60~K remains in the solid state, as is found in the infrared experiments. This solid-state CO$_2$ is released into the gas phase during TPD due to thermal desorption, and subsequently measured with the QMS. However, during VUV irradiation there is some gas-phase CO$_2$ signal measured by the QMS. This is illustrated in Fig.~\ref{fig:exp_40K}c (shaded areas) by the signal at mass-to-charge ratio ($m/z$)~=~49, which is associated with the main peak of the $^{13}$C$^{18}$O$_2$ mass spectrum. The increase in this CO$_2$ QMS signal follows approximately the same trend as the growth of the CO$_2$ column density measured in the infrared (Fig.~\ref{fig:exp_40K}a). When the VUV shutter is closed, non-shaded areas, the signal at $m/z$~=~49 drops. We attribute this gas-phase CO$_2$ QMS signal to photodesorbed CO$_2$ from the solid state \citep{2014_Fillion_CO2_PD_FaDi..168..533F}.

During TPD, there are two distinct desorption peaks of CO$_2$ with an elevated plateau between them (see e.g., Fig.~\ref{fig:exp_40K}d). The first desorption peak occurs at 78~K, the canonical desorption temperature of CO$_2$. The CO$_2$ molecules that desorb at this temperature, are those in CO$_2$ aggregates. The second desorption peak coincides with the water desorption peak observed at 156~K. Both behave fully in agreement with the deconvolved infrared components at 2279 and 2243~cm$^{-1}$ (see e.g., Fig.~\ref{fig:exp_40K}b).

\subsubsection{Gas-phase CO$_2$ ($\geq70$~K)}
In the remainder of the experiments, listed in Table~\ref{tab:exp}, and shown in Figs.~\ref{fig:exp_70K}, \ref{fig:exp_80K}--\ref{fig:exp_120K}, the temperature of the ASW ranges from 70--120~K. In the temperature range 70--90~K equal amounts of CO are converted into CO$_2$ as compared to < 70~K. However, the majority of the formed CO$_2$ is immediately released into the gas phase after formation. Similar to Fig.~\ref{fig:exp_40K}, we present the results of ASW at 70~K in Fig.~\ref{fig:exp_70K}. During this experiment $\sim20\%$ of the formed CO$_2$ stays on the surface of ASW, while the remainder is released into the gas phase. The release of CO$_2$ into the gas phase is slightly below the canonical CO$_2$ desorption temperature, 78~K. This is no surprise as the binding energy of CO$_2$ bound to H$_2$O equals 2250~K, while the binding energy between CO$_2$ molecules is higher at 2415~K \citep{2017_Jiao_CO2_ApJ...837...65H}. During the first hour of VUV irradiation, the ASW surface ``traps'' some of the formed CO$_2$ in its deep binding sites, but once these are occupied, most of the subsequently formed CO$_2$ is released into the gas phase. This is reflected by the initial rapid build up of CO$_2$ in the infrared during the first VUV irradiation interval (Fig.~\ref{fig:exp_70K}a). Additionally, in Fig.~\ref{fig:exp_70K}c it is shown that the gas-phase CO$_2$ builds up during the initial VUV interval, where it reaches steady state at the same time when the growth of solid-state CO$_2$ levels off. Lastly, during TPD the two main desorption features appear at approximately 80 and 155~K, but not as prominent as in the 40--60~K experiments. The majority of the CO$_2$ is released during TPD in the ``plateau'' region between 85--145~K, that is, between the canonical desorption of CO$_2$ aggregates and H$_2$O, as is shown in Fig.~\ref{fig:exp_70K}d.

In the experiments with ASW temperatures between 90 and 120~K (Figs.~\ref{fig:exp_90K}--\ref{fig:exp_120K}), no solid-state CO$_2$ is observed in the infrared (column density $\leq 2.0\times10^{13}$~cm$^{-2}$). The TPDs in this temperature range, however, do reveal that some CO$_2$ is still bound to the surface of the ASW. Following the trend as seen in the experiments with ASW at 70 and 80~K, the amount of CO$_2$ that remains on the ASW surface decreases with increasing temperature, see Figs.~\ref{fig:exp_90K}--\ref{fig:exp_120K}. This is in line with the decrease in absolute signal of the ``plateau'' during TPD. Interestingly, a significant amount of O$_2$ formation is observed in the 90--120~K temperature range, as shown by the QMS signal at $m/z$~=~36 representing $^{18}$O$_2$. The formation of O$_2$ increases with temperature at the cost of CO$_2$. At 120~K the formation of CO$_2$ is almost completely quenched. Compared to 80~K the raw QMS data at 120~K for CO$_2$ is decreased by over a factor of 10 and the O$_2$ signal increased by over a factor of 50.

\subsection{Control experiments}
Several control experiments have been performed in order to aid in the investigation of the interaction between gas-phase CO and VUV irradiated water ice. Specifically, a control experiment where a different oxygen isotope is used in CO, an experiment without VUV irradiation, and an experiment where gas-phase CO is omitted, as listed in Table~\ref{tab:exp}. The results of these experiments are presented in Appendix~\ref{fig:exp_isotope}, \ref{fig:exp_no_uv}, and \ref{fig:exp_water_only}, respectively.

\begin{figure}[t!]
    \includegraphics[width=\hsize]{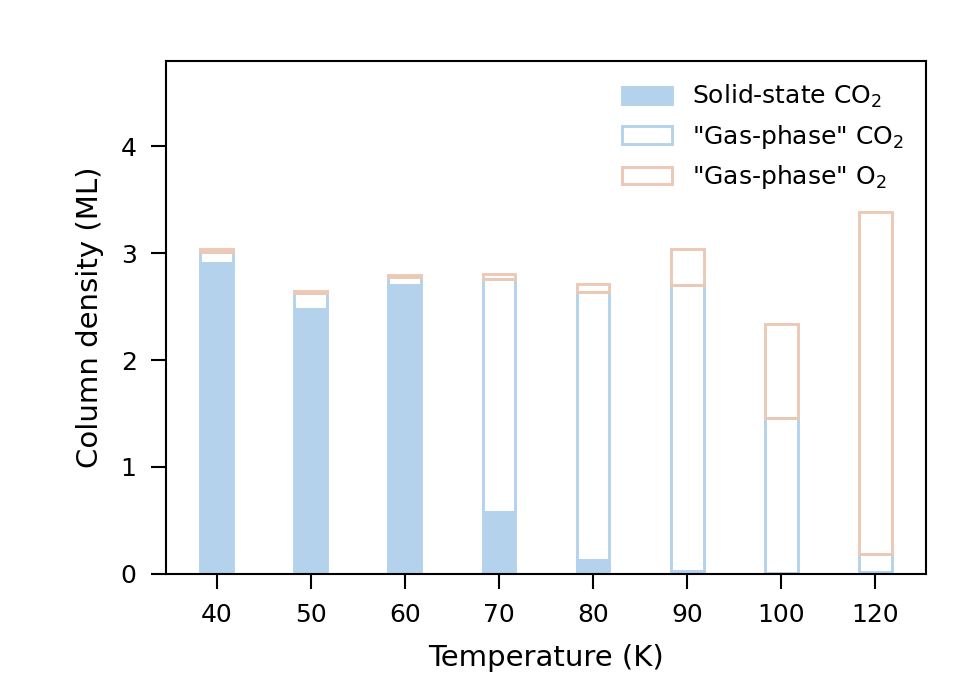}
    \caption{Column densities of the formed products when ASW is VUV irradiated and exposed to gas-phase CO as function of temperature. Solid-state column densities are derived through RAIRS and gas-phase column densities through the QMS as described in Sect.~\ref{sec:analysis}. Both the RAIRS and gas-phase data provide information on the gas-surface interactions studied here.} \label{fig:tot_col_den}
\end{figure}

The experiment with gas-phase $^{13}$C$^{16}$O, instead of the $^{13}$C$^{18}$O isotopologue, provides additional information on the interaction between the gas-phase CO and VUV irradiated water ice (Fig~\ref{fig:exp_isotope}). In the infrared the three main absorption features are shift by approximately 20~cm$^{-1}$, 2297, 2280, and 2262~cm$^{-1}$, with respect to the main experiments, 2279, 2260, and 2243~cm$^{-1}$, respectively. This indicates that these three CO$_2$ features are formed through the same process as each of them shift by approximately the same wavenumber due to the difference in mass of the oxygen isotope. Additionally, the QMS results show that the CO$_2$ is now detected at a $m/z$ that equals 47 ($^{13}$C$^{16}$O$^{18}$O) instead of 49 ($^{13}$C$^{18}$O$_2$). This indicates that both oxygen atoms from the precursors, that is, $^{16}$O from $^{13}$C$^{16}$O and $^{18}$O from H$_2$$^{18}$O, are involved in the formation of CO$_2$.

The experiment without VUV irradiation, shown in Fig~\ref{fig:exp_no_uv}, is performed to track the level of CO$_2$ from other sources. A build up of~0.1 ML CO$_2$ is detected during the time that the water ice is exposed to gas-phase CO. Compared to the counterpart experiment with VUV radiation, shown in Fig~\ref{fig:exp_40K}, this is only 4\% of the total CO$_2$ formed when water ice is VUV irradiated. This small contribution is most likely due to contamination from previous experiments or trace amounts of $^{13}$C$^{18}$O$_2$ in our $^{13}$C$^{18}$O gas bottle. Regardless, this amount of contamination is negligible.

Lastly, water ice is VUV irradiated without the presence of gas-phase CO in the vacuum chamber. The results of this experiment are presented in Fig~\ref{fig:exp_water_only}. Both the infrared and the QMS show at most trace amounts of CO$_2$. This supports that the main processes through which CO$_2$ in this study is formed through the interaction between gas-phase CO and VUV irradiated water ice. Additionally, the QMS shows the release of O$_2$ into the gas phase during VUV irradiation, which is not seen in the counterpart main experiment with gas-phase CO, as shown in Fig~\ref{fig:exp_40K}.

\subsection{CO$_2$ and O$_2$ column densities} \label{ssec:col_dens}
For each of the main experiments the column densities of the products, with a total VUV incident fluence of $4.5 \times 10^{18}$~photons, are summarized in Fig.~\ref{fig:tot_col_den}. The column density of solid-state CO$_2$ is derived through the combined integrated absorbance area of the three infrared CO$_2$ features. The CO$_2$ and O$_2$ gas-phase column densities are derived through the calibration of the QMS described in Sect.~\ref{ssec:QMS_cal}. In short, in the temperature range 40--60~K the main product is solid-state CO$_2$. As substrate temperatures at 70~K or above, the CO$_2$ is detected in the gas phase, and at even higher temperatures, that is, $\geq90$~K, O$_2$ formation is observed at the cost of CO$_2$ production.

\section{Discussion} \label{sec:disc}
It is clear from the presented results that CO$_2$ is formed in our experiments, and that the temperature of the ASW influences the physical appearance of CO$_2$. In the following section we explore the different pathways to CO$_2$, which pathway results in the formation of CO$_2$ in our experiments, and the CO$_2$ production efficiency per absorbed VUV photon.

\subsection{Exploring the reaction network} \label{ssec:network}
In the introduction, we mentioned several solid state pathways that can form CO$_2$. The formation of CO$_2$ in our experiments is driven by VUV irradiation of ASW that interacts with gas-phase CO. This is different from most earlier studies where CO was embedded and intimately mixed with water ice. Such experiments are relevant for astronomical scenarios in which H$_2$O and CO are mixed in the solid state. However, these conditions are different from those discussed later in this study, that is, protoplanetary disks (Sect.~\ref{ssec:ppds}) and at molecular cloud edges (Sect.~\ref{ssec:mol_clouds}).

\begin{figure}[ht!]
\centering
    \includegraphics[width=0.99\hsize]{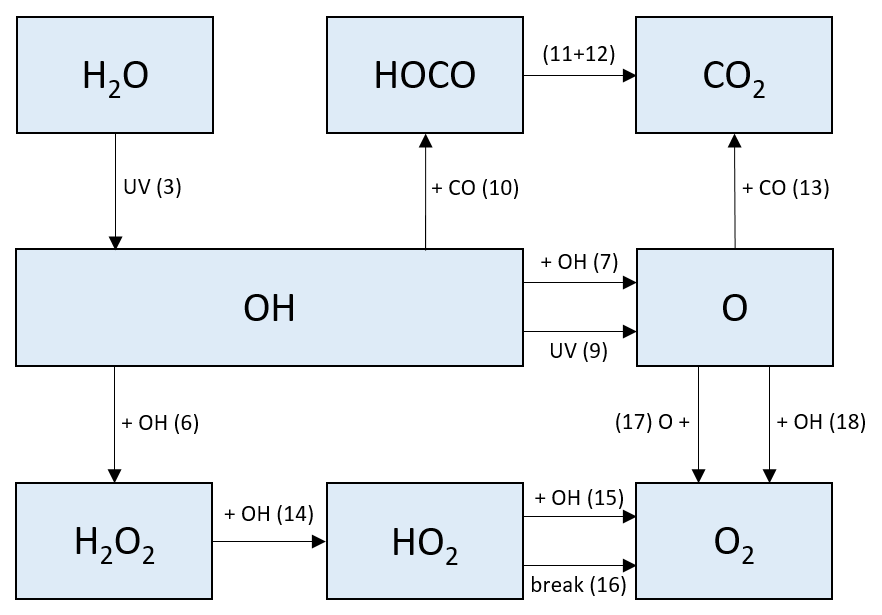}
    \caption{Schematic overview of the reactions that potentially occur in our experiments with VUV irradiation of solid-state H$_2$O in the presence of gas-phase CO. The numbers in parenthesis correspond to the reactions in Sect.~\ref{ssec:network}. Backward, recombination, and by-products of reactions are not shown for clarity.} \label{fig:network}
\end{figure}

There are two potential pathways to form CO$_2$ in our experiments, which involve both H$_2$O and CO, and three possible pathways that could lead to the observed formation of O$_2$ at higher temperatures. Figure~\ref{fig:network} gives a schematic overview of these reactions. In general, UV photons dissociate H$_2$O in the solid state through its excited \~{A} and \~{B} states, which mainly lead to the formation of OH radicals and atomic oxygen,
\begin{linenomath}
\begin{equation} \label{rea:H_OH}
    \mathrm{H_2O} + h\nu \rightarrow \mathrm{H + OH},
\end{equation}
\begin{equation} \label{rea:H2_O1d}
    \mathrm{H_2O} + h\nu \rightarrow \mathrm{H_2 + O}.
\end{equation}
\end{linenomath}
\citet{1975_Stief_O1D_JChPh..62.4000S} reported quantum efficiencies for both reactions~(\ref{rea:H_OH})~and~(\ref{rea:H2_O1d}) upon irradiation in two different wavelength ranges, namely 145--185~nm and 105--145~nm corresponding with the excited \~{A} and \~{B} states of solid-state H$_2$O, respectively. Water dissociated through the excited \~{A} state has been reported to have quantum efficiencies of 0.99 and $\leq0.01$ for reactions~(\ref{rea:H_OH})~and~(\ref{rea:H2_O1d}), respectively. The dissociation through the excited \~{B} state was reported to have quantum efficiencies of 0.89 and 0.11, respectively.

In this study we use a MgF$_2$ window with a cut-off wavelength above the wavelength of Lyman-$\alpha$ photons. This ensures that Lyman-$\alpha$ photons from the MDHL are absorbed, and that the majority of the UV photons are in the 140--170~nm (7.3--8.9~eV) range, see Fig.~\ref{fig:uv_spec}. This, combined with the reported quantum efficiencies, results in the dissociation of H$_2$O only through its excited \~{A} state, producing mainly OH radicals through reaction~(\ref{rea:H_OH}). The removal of Lyman-$\alpha$ photons makes the VUV spectrum less representative of those in interstellar environments. However, it does allow for an in-depth investigation of primary reactions including only OH radicals.

The formed H and OH proceed in different ways depending on the depth in the ice at which dissociation occurs. The molecular dynamics calculations by \citet{2008_Andersson_H2O_MD_A&A...491..907A} showed that in the top three monolayers the majority of the photodissocation events results in the desorption of H and trapping of OH. At four monolayers or deeper most of the photodissocation events result in trapping of both species or recombination, reforming H$_2$O through reaction~(\ref{rea:H2O}),
\begin{linenomath}
\begin{equation} \label{rea:H2O}
 \mathrm{H + OH \rightarrow H_2O}.
\end{equation}
\end{linenomath}
The desorption of H in the top three monolayers results in an enrichment of OH radicals on the surface. It was found in these calculations that the OH radicals can diffuse up to 60 \r{A} on top of the H$_2$O surface at 10~K. This diffusion occurs on picosecond timescales, and does not include any thermal diffusion on longer timescales. \citet{2009_Hama_OH_desorp_JChPh.131e4508H} showed that the OH radicals produced through reaction (\ref{rea:H_OH}) are hot and have a translational temperature of $1300\pm300$~K. This significant amount of translational energy allows for additional diffusion, and increases the probability of two OH radicals to meet and react with each other. This reaction either forms hydrogen peroxide (H$_2$O$_2$) or H$_2$O and atomic oxygen, see reactions (\ref{rea:H2O2}) and (\ref{rea:H2O_O3p}), respectively. The branching ratio between reactions~(\ref{rea:H2O2}) and (\ref{rea:H2O_O3p}) was found to be 0.8 and 0.2 for two nonenergetic OH radicals reacting with each other at 40--60~K \citep{2011_Oba_CO_OH_40-60K_PCCP...1315792O},
\begin{linenomath}
\begin{equation} \label{rea:H2O2}
 \mathrm{OH + OH \rightarrow H_2O_2},
\end{equation}
\begin{equation} \label{rea:H2O_O3p}
 \mathrm{OH + OH \rightarrow H_2O + O}.
\end{equation}
\end{linenomath}
As the abundance of the H$_2$O$_2$ increases, also the amount of H$_2$O$_2$ dissociated by VUV photons increases through reaction (\ref{rea:OH_OH}),
\begin{linenomath}
\begin{equation} \label{rea:OH_OH}
    \mathrm{H_2O_2} + h\nu \rightarrow \mathrm{OH + OH}.
\end{equation}
\end{linenomath}
The translational temperature of these OH radicals was found to be $7500\pm1000$~K, which potentially allows for even further diffusion of OH radicals across the surface \citep{2009_Hama_OH_desorp_JChPh.131e4508H}. These OH radicals on the surface, or those below for that matter, can be subsequently dissociated by VUV photons forming atomic hydrogen and oxygen, see reaction (\ref{rea:H_O3p}),
\begin{linenomath}
\begin{equation} \label{rea:H_O3p}
    \mathrm{OH} + h\nu \rightarrow \mathrm{H + O}.
\end{equation}
\end{linenomath}

The OH radicals formed through reactions~(\ref{rea:H_OH}) and (\ref{rea:OH_OH}) are potential candidates for CO$_2$ formation when reacting with gas-phase CO. In the experimental study by \citet{2010_Oba_non_OH_CO_ApJ...712L.174O} the formation of CO$_2$ was observed from co-deposition of nonenergetic OH radicals, cooled to 100~K prior to deposition, and CO molecules at 10 and 20~K. The authors proposed that CO$_2$ forms through reactions~(\ref{rea:trans-HOCO}), (\ref{rea:cis-HOCO}), and (\ref{rea:CO2_H}),
\begin{linenomath}
\begin{equation} \label{rea:trans-HOCO}
    \mathrm{CO + OH \rightarrow }\ trans\mathrm{-HOCO},
\end{equation}
\begin{equation} \label{rea:cis-HOCO}
     trans\mathrm{-HOCO \rightarrow}\ cis\mathrm{-HOCO},
\end{equation}
\begin{equation} \label{rea:CO2_H}
    cis\mathrm{-HOCO \rightarrow CO_2 + H}.
\end{equation}
\end{linenomath}
Atomic oxygen formed through reactions~(\ref{rea:H2O_O3p}) and (\ref{rea:H_O3p}) also has the potential to react with CO and form CO$_2$ through reaction~(\ref{rea:CO2}),
\begin{linenomath}
\begin{equation} \label{rea:CO2}
    \mathrm{CO + O \rightarrow CO_2}.
\end{equation}
\end{linenomath}
The addition of atomic oxygen to CO has been experimentally shown to work between 5 and 20~K, where CO is adsorbed on a bare substrate \citep{2001_Roser_CO+O_ApJ...555L..61R, 2006_Madzunkov_CO+O_PhRvA..73b0901M, 2011_Raut_CO_O_ApJ...737L..14R, 2013_Ioppolo_RScI...84g3112I}. The same reaction has also been investigated on top of ASW by \citet{2013_Minissale_CO_O_A&A...559A..49M}. These authors show that the CO$_2$ is formed through reaction~(\ref{rea:CO2}) when CO and O are co-deposited on the surface of ASW in the temperature range 10--50~K.

The initially formed O, OH, and H$_2$O$_2$ can react with each other to form the hydroperoxyl radical (HO$_2$) and O$_2$. The HO$_2$ radical is formed through subsequent reactions of H$_2$O$_2$ with OH, see reaction~(\ref{rea:H2O2_OH}). Molecular oxygen can be formed through different means, a) an HO$_2$ radical reacts with OH, see reaction~(\ref{rea:HO2_OH}), b) the HO$_2$ radical falls apart, see reaction~(\ref{rea:HO2_dism}), c) atomic oxygen reacts with another atomic oxygen, see reaction~(\ref{rea:O_O}), or d) atomic oxygen reacts with OH, see reaction~(\ref{rea:O_OH}).
\begin{linenomath}
\begin{equation} \label{rea:H2O2_OH}
    \mathrm{H_2O_2 + OH \rightarrow HO_2 + H_2O},
\end{equation}
\begin{equation} \label{rea:HO2_OH}
    \mathrm{HO_2 + OH \rightarrow O_2 + H_2O},
\end{equation}
\begin{equation} \label{rea:HO2_dism}
    \mathrm{HO_2 \rightarrow H + O_2},
\end{equation}
\begin{equation} \label{rea:O_O}
    \mathrm{O + O \rightarrow O_2},
\end{equation}
\begin{equation} \label{rea:O_OH}
    \mathrm{O + OH \rightarrow O_2 + H}.
\end{equation}
\end{linenomath}

\subsection{CO$_2$ formation pathway} \label{ssec:pathway}
In order to disentangle which of the above formation pathways is active in our experiments, we look into the temperature dependence of the CO$_2$ formation process and which isotopes are incorporated in the produced CO$_2$. We also extensively compare with literature data. Across the temperature range of 40--90~K, the total column density of formed CO$_2$ is constant, but above 90~K the efficiency of CO$_2$ formation decreases due to the competing formation of O$_2$, see Fig.~\ref{fig:tot_col_den}. This temperature dependence contains significant amount of information, which allows us to constrain the formation of CO$_2$ to one pathway. 

Our experiments show that CO$_2$ is formed through the reaction between gas-phase CO and solid-state OH radicals. These OH radicals are the primary product of H$_2$O dissociation, and are thus most likely to react with CO. This particular reaction pathway to CO$_2$ has been investigated extensively \citep{2002_Watanabe_CO2_ApJ...567..651W, 2007_Watanabe_H2O-CO_ApJ...668.1001W, 2009_Ioppolo_CO2_A&A...493.1017I, 2010_Oba_non_OH_CO_ApJ...712L.174O, 2011_Oba_CO_OH_40-60K_PCCP...1315792O, 2011_Ioppolo_CO2_MNRAS.413.2281I, 2011_Noble_CO2_ApJ...735..121N, 2011_Zins_CO+OH_ApJ...738..175Z, 2014_Yuan_ERCO2_ApJ...791L..21Y}. However, the majority of these studies were performed at temperatures where CO is in the solid state, and mixed with H$_2$O. \citet{2010_Oba_non_OH_CO_ApJ...712L.174O} looked at the formation of CO$_2$ through co-deposition of CO and nonenergetic OH radicals at a temperature of 10 and 20~K. Besides CO$_2$, the authors also observed the intermediate products cis- and trans-HOCO radicals at 1774 and 1812~cm$^{-1}$, respectively. They found that the HOCO absorption features disappear at $T$ > 40~K, which is in line with the experimental work of \citet{1971_Milligan_CO_OH_JChPh..54..927M}. This is most likely the reason why the cis- and trans-HOCO radicals are not detected in the infrared spectra of our experiments (spectra not shown). In a follow-up study, \citet{2011_Oba_CO_OH_40-60K_PCCP...1315792O} investigated the same reactions, but in the temperature range 40--60~K. Formation of CO$_2$ was observed in the infrared, but the efficiency at which CO was converted into CO$_2$ decreased with increasing temperature. The conversion rates were found to be 1.4\%, 0.8\%, and 0.3\% at 40, 50, and 60~K, respectively. This decrease was attributed to the decreasing residence times with increasing surface temperature of both CO and OH. In our experiments this efficiency decrease is not observed, even when only considering solid-state CO$_2$. A possible explanation for this is the different origin of the OH radicals; in our work the radicals are formed in situ with excess energy, whereas in previous studies OH radicals were deposited.

The other proposed formation pathway to CO$_2$, that is, atomic oxygen reacting with CO, can be excluded. This is because atomic oxygen can only originate in our experiments as a secondary product through reactions~(\ref{rea:H2O_O3p}) and (\ref{rea:H_O3p}). Additionally, on ASW atomic oxygen and CO have similar binding energies, that is, 1320 and 1350~K, respectively, and thus their residence times on the surface are comparable \citep{2016_minissale_O_N_A&A...585A.146M, 2016_Jiao_bind_energy_ApJ...825...89H}. Because of their similar residence times, no difference should be observed between the formation of CO$_2$ and O$_2$ with experimental temperature. However, O$_2$ is only significantly formed at temperatures $\geq90$~K. This is proof that atomic oxygen is not involved in the formation of CO$_2$. Lastly, \citet{2013_Minissale_CO_O_A&A...559A..49M} investigated the formation of CO$_2$ through co-deposition of CO and atomic oxygen on top of ASW. It was found that the efficiency of CO$_2$ formation peaked at 35~K and dropped to zero at 60~K. As CO$_2$ formation is observed in our experiments up to 120~K, this is again evidence that atomic oxygen is not involved in the formation of CO$_2$ in our experiments. It should be noted that experimental conditions are not identical, as in our experiments the atomic oxygen would be formed in situ instead of co-deposited with CO. However, since atomic oxygen would be in the ground state, for both our work and that of \citet{2013_Minissale_CO_O_A&A...559A..49M}, no clear differences are expected.

Formation of CO$_2$ through excited CO* reacting with another CO molecule on the surface of ASW can also be ruled out. This is unlikely to occur, because it would require a CO molecule to be excited during its short, but nonzero, residence time and react with another CO molecule which has an equally short residence time. Additionally, in the control experiment with H$_2${}$^{18}$O and $^{13}$C$^{16}$O, the formed CO$_2$ is measured with the QMS during TPD at $m/z$~=~47, corresponding to $^{13}$C$^{16}$O$^{18}$O, see Appendix~\ref{fig:exp_isotope}. If the CO$_2$ would be formed through excited CO and another CO molecule it would be expected to be detected at $m/z$~=~45, corresponding to $^{13}$C$^{16}$O$_2$.

From above results and discussion it is most likely that CO$_2$ is formed through the interaction between CO and OH radicals, formed by UV dissociation of H$_2$O. However, it is not yet clear if CO directly interacts with OH radicals from the gas phase, that is, an Eley-Rideal type reaction, or if CO adsorbs onto the ASW, diffuses, and subsequently reacts with OH radicals, that is, a Langmuir-Hinshelwood type reaction. Additionally, the formation location of CO$_2$ is also not yet clear; is it formed on the surface or embedded in the ASW? Both of these topics are discussed in the following section.

\subsection{Formation location of CO$_2$} \label{ssec:location}
The average time a species resides on a surface can be estimated at a given temperature starting from the measured or calculated binding energy to that surface. We derive that CO has residence times on ASW of $4.5\times10^{2}$--$7.7\times10^{-8}$~s in the range from 40 to 120~K. This is found through the Arrhenius equation, which can be written as
\begin{linenomath}
\begin{equation} \label{eq:arrhenius}
    k = A e^{-\frac{E_{bind}}{T}},
\end{equation}
\end{linenomath}
where $k$ is the rate constant, $A$ the frequency factor, which is taken to be 10$^{12}$ s$^{-1}$, $E_{bind}$ the binding energy of a species to a specific surface in K, and $T$ the temperature of the surface in K. The residence time is then given by the reciprocal of the rate constant from Eq.~\ref{eq:arrhenius}. For CO on ASW, the binding energy is dependent on the CO surface coverage, ranging from 1000--1700~K at 1--10$^{-3}$~monolayer coverage \citep{2016_Jiao_bind_energy_ApJ...825...89H}. The above residence times are estimated given an average binding energy of 1350~K for CO on ASW. For comparison, the binding energy of CO on the CO--CO interface is $855\pm25$~K \citep{2005_Oberg_CO_bind_ApJ...621L..33O}.

Even within these short residence times, some diffusion across the surface is expected. The number of binding sites CO visits on ASW, before desorption occurs, is estimated to be $7.2\times10^{10}$--$4.2\times10^{3}$ in the range 40--120~K. This is also derived through Eq.~\ref{eq:arrhenius}. Specifically, the number of different binding sites a molecule can visit before a species desorbs is approximated by dividing the diffusion rate by the desorption rate. The diffusion rate is estimated by exchanging the $E_{bind}$ term in Eq.~\ref{eq:arrhenius} for the diffusion energy ($E_{diff}$). The rate constant is then a proxy of the number of hops a species makes between different binding sites per second. The diffusion energy for CO on ASW has recently been measured in situ with transmission electron microscopy (TEM), and was found to be $350\pm50$~K \citep{2020_Kouchi_Ediff_CO_CO2_ASW_ApJ...891L..22K}. 

Given these residence times and amount of binding sites that are ``visited'' before desorption occurs, we conclude that CO spends sufficient time on the surface of ASW to react with OH radicals through a Langmuir-Hinshelwood type reaction. This is different from \citet{2014_Yuan_ERCO2_ApJ...791L..21Y}, who investigated this reaction under similar experimental conditions and attributed it to an Eley-Rideal type of reaction. \citet{2014_Yuan_ERCO2_ApJ...791L..21Y} employed a slightly higher binding energy of 0.125~eV (1450~K) for CO on H$_2$O, which results in a residence time of $\sim 2 \times 10^{-4}$~s at their experimental temperature of 76~K. The residence time is used to calculate the fractional coverage of CO on H$_2$O and was found to be $1 \times 10^{-6}$~ML. The resulting fractional coverage of OH was derived to be 0.05~ML, over four orders of magnitude higher, which led to the conclusion of an Eley-Rideal type of reaction. However, diffusion of CO during this (short) residence time was not considered. At these temperatures, CO visits approximately $10^6$ binding sites during its residence time, and thus, the effective surface scanned by CO is $\sim1$~ML even though the fractional coverage of CO is only $1 \times 10^{-6}$~ML. This supports that the involved mechanism follows a Langmuir-Hinshelwood type reaction.

Furthermore, we see no evidence of significant CO diffusion, and subsequent trapping, into the bulk of the H$_2$O ice. However, there is some trapping of CO on the surface or pores of the ASW. This is shown in a control experiment where ASW is exposed to CO molecules, but not to VUV irradiation. During TPD of this control experiment, as is shown in Appendix~\ref{fig:exp_no_uv}, the majority of the CO desorbs at approximately 50~K. Only a small amount of CO ``volcano'' desorbs when the ASW crystallizes. It is most likely that this CO got trapped in ASW due to pore collapse, instead of actually diffusing into the bulk ASW. 

\subsection{Temperature dependent formation, CO$_2$ vs O$_2$} \label{ssec:CO2vsO2}
At ASW temperatures above 90~K, the production of O$_2$ increases at the cost of CO$_2$ formation (see e.g., Fig~\ref{fig:tot_col_den}). An in-depth investigation of O$_2$ formation is beyond the scope of this work, but it is briefly discussed to explain the decrease in CO$_2$ production. For more information on O$_2$ production, we refer to a recent study that quantitatively investigated the production of O$_2$ and H$_2$O$_2$ by VUV irradiation of H$_2$O \citep{2022_Bulak_O2_A&A...657A.120B}.

Since O$_2$ formation occurs at the cost of CO$_2$ production, it is likely that both species have a common precursor. In Sect~\ref{ssec:network}, the pathways to O$_2$ are through the OH radical or atomic oxygen. Due to the high temperature and simultaneous increase and decrease in O$_2$ and CO$_2$ abundance, respectively, the O$_2$ is likely formed through OH radicals. These OH radicals form O$_2$ sequentially through H$_2$O$_2$ and HO$_2$, see reactions~(\ref{rea:H2O2_OH}), (\ref{rea:HO2_OH}), and (\ref{rea:HO2_dism}). The formation of O$_2$ involving atomic oxygen is unlikely, as it is shown to not be involved in the formation of CO$_2$, see Sect.~\ref{ssec:pathway}. Additionally, in our experiments atomic oxygen is a secondary product through reactions~(\ref{rea:H2O_O3p}) and (\ref{rea:H_O3p}). The low residence times of this atomic oxygen, $<2.3\times10^{-6}$~s at temperatures above 90~K, combined with low availability makes formation of O$_2$ through OH radicals the dominant pathway.

The formation of CO$_2$ and O$_2$ is dependent on the availability of OH radicals and the temperature of the ASW. Given this temperature dependence, it is likely that the increased mobility of the OH radical and reduced residence time of CO at high temperatures holds the answer to why O$_2$ formation is more favourable. The pathway to form O$_2$ at high temperatures is worthy of further investigation.

\subsection{Conversion rate of CO into CO$_2$} \label{ssec:CO_into_CO2}
In order to demonstrate that the gas-grain pathway to convert CO into CO$_2$ is a process of importance in astrophysical environments, we discuss in this section the conversion rate and limiting factors in our experiments. In total $\sim60$~monolayers of water ice are deposited on the substrate in preparation of our experiments. However, as this conversion of CO into CO$_2$ occurs on the surface, not all of this H$_2$O is available to act as a reacting medium. Classically, the surface of solid-state H$_2$O contains approximately $10^{15}$~molecules per cm$^2$. However, due to the porous nature of our ASW, the available H$_2$O surface for CO to adsorb on is expected to be larger. Additionally, with hydrogen released from the top three monolayers upon UV dissociation of H$_2$O, and the mobility of the OH radicals, we assume that OH radicals formed in the top three monolayers are available to convert CO into CO$_2$ \citep{2008_Andersson_H2O_MD_A&A...491..907A}. These top three monolayers, that is, $3.0\times10^{15}$~H$_2$O molecules cm$^{-2}$, are henceforth the reactive surface.

The VUV radiation that impacts the water ice, consists only of photons from molecular H$_2$ emission, as the Lyman-$\alpha$ photons of the MDHL are absorbed by our MgF$_2$ window, see Fig.~\ref{fig:uv_spec}. The average H$_2$O absorption cross section for the VUV photons is taken to be ($1.2\pm0.1$)~$\times10^{-18}$~cm$^2$. This average is derived from the absorption cross section of the two main molecular H$_2$ emission peaks at 157.8 and 160.8~nm in a one-to-one ratio \citep{2014_Cruz-Diaz_VUV_cross_A&A...562A.119C}. Given this cross section and the Beer-Lambert law, $\sim0.4\%$ of the total incident VUV fluence is absorbed by H$_2$O in the reactive surface, and this equals $1.6\times10^{16}$ photons. On the assumption that H$_2$O dissociation is 100\% efficient, this produces an equal amount of OH radicals in the reactive surface. In the temperature range 40--90 K, approximately $2.7\times10^{15}$~CO$_2$ molecules are formed, and thus an equal amount of OH radicals is consumed through reactions (\ref{rea:trans-HOCO}--\ref{rea:CO2_H}). This means that, for the above assumption, only 17\% of the available $1.6\times10^{16}$~OH radicals are involved in the conversion of CO into CO$_2$. The question is, what is the limiting factor that determines this efficiency factor?

In our experiments, it is the amount of OH and not CO that is limiting the formation of CO$_2$. In the temperature range of 40--90 K, the formation of CO$_2$ is considered to be constant. However, the residence time of CO is lowered by a factor of $10^{8}$ and the binding sites visited by CO by a factor of $10^5$ from 40 to 90~K. That is to say, once CO finds an OH radical on the surface that is available, the conversion into CO$_2$ is (close to) unity. The relatively small fraction (17\%) of VUV absorption events resulting in the CO$_2$ production may be due to a smaller number of available OH radicals, for example, the recombination probability of H + OH on the surface is high. It is also possible that other products are formed besides CO$_2$ or that UV dissociation of ASW is not unity, which is typically assumed. These are in our view the most logical explanations. Below we argue why from these options the dissociation efficiency of solid-state H$_2$O is most likely the limiting factor in the formation of CO$_2$.

It would be surprising if a significant portion of the available OH radicals would not react, because even nonenergetic OH radicals in the ground state are able to form CO$_2$ with CO at 10~K \citep{2010_Oba_non_OH_CO_ApJ...712L.174O}. Formation of other products, such as, H$_2$O$_2$, is also excluded as we do not detect significant amounts of the possible products in the experiments. Assuming that VUV photodissociation of water ice is unity and recombination does not occur in the reactive surface, the reactive surface would be dominated by OH radicals (83\% of the remaining VUV absorption events). This amount of OH radicals should largely find each other and react to form H$_2$O$_2$ and O$_2$. The infrared does not show any of the vibrational modes of H$_2$O$_2$ within our detection limits (spectra not shown), especially not with the expected H$_2$O$_2$ column density of $\sim7$~monolayers, assuming that all of the remaining OH radicals react with each other and form H$_2$O$_2$ through reaction (\ref{rea:H2O2}). In itself this column density is already questionable, as it is approximately twice the amount of available H$_2$O molecules in the reactive surface. Additionally, a control experiment where ASW is irradiated at 40~K and gas-phase CO is omitted does show the formation of H$_2$O$_2$ and O$_2$, see Appendix \ref{fig:exp_water_only}. This shows that in the presence of gas-phase CO the formation of H$_2$O$_2$ and O$_2$ in the reactive surface is quenched, and that in the main experiments all available OH radicals react with CO into CO$_2$.

The amount of CO$_2$ produced, and lack of other products, in the experiments points at that either the recombination of H + OH is high in the reactive surface or that not every VUV absorption event results in the dissociation of H$_2$O. Fully investigating this question is beyond the scope of this work, but it leads to a conundrum that requires attention. It is unlikely that recombination to H$_2$O is significantly more efficient than predicted by molecular dynamics calculations. Photodissociation and desorption occur on picosecond timescales after VUV absorption in these simulations, and thus, diffusion of atomic hydrogen within this time window is improbable. It should be noted that in molecular dynamics calculations only a single event is considered per simulation, and thus for H$_2$O recombination to occur the atomic hydrogen needs to find its original OH partner before it desorbs. It could be that in our experiments the atomic hydrogen reforms H$_2$O with previously formed OH radicals due to the high VUV photon fluxes, that is, $2.5\times10^{14}$~photons s$^{-1}$ cm$^{-2}$. Molecular dynamics calculations of H$_2$O photodissociation and recombination with neighbouring available OH radicals are needed to test if this pathway is viable. For now we deem it inefficient due to the extremely short time scales in which desorption occurs after dissociation.

A lower dissociation efficiency of water ice upon absorption of a VUV photon, that is, well below unity, seems to be the most likely explanation for our findings, at least based on the processes discussed here. In Fig.~2 of \citet{2008_Andersson_H2O_MD_A&A...491..907A} the fractional probabilities of photodissociation pathways for H$_2$O are given per absorbed photon and sum to (near) unity. However, per design these simulations only follow H$_2$O that is photodissociated. \citet{1990_Schriever_dissociation_H2O_JChPh..93.9206S} investigated the absolute photodissociation quantum yield of H$_2$O in an argon matrix (ratio 1:500). It was found that at 160~nm and 5~K the photodissociation efficiency of isolated H$_2$O in argon equals 20--30\%. It should be noted that in their work the H$_2$O is isolated and trapped in the bulk argon, which cannot be extrapolated to our work. That being said, it does match well with our 17\% efficiency, which suggests that indeed the dissociation of water ice, through the excited \~{A} state, is not unity. Similar results have been found by \citet{2018_Kalvans_eff_dissociation_MNRAS.478.2753K} who investigated how the photodissociation efficiency differs in general for a molecule in the gas-phase and solid-state. They found a best-fit value of 0.3 for the ratio solid-state to gas-phase photodissociation from their 1D astrochemical model in comparison to line-of-sight observations of collapsing interstellar clouds. Even though our findings are not fully conclusive, it seems to be in line with results presented earlier in the literature.

\citet{2013_Arasa_HOCO_JPCA..117.7064A} investigated the CO + OH pathway at 10~K with molecular dynamics calculations and found a conversion probability for CO$_2$ of ($3.6\pm0.7$) $\times10^{-4}$ per absorbed photon, which is significantly lower than our 17\%. However, it was found that the formation probability of the intermediate HOCO complex is two orders of magnitude higher with ($3.00\pm0.07)$ $\times10^{-2}$ per absorbed photon. This is explained by the HOCO complex being trapped in the solid state and losing its internal energy to the surrounding molecules, which prevents further reaction to CO$_2$ + H. This does not necessarily align with the findings presented here. In our experiments the HOCO complex is not observed. However, it could very well be that this HOCO complex is briefly present, but transfers into CO$_2$ due to the increased ASW temperature, as seen in \citet{1971_Milligan_CO_OH_JChPh..54..927M} and \citet{2011_Oba_CO_OH_40-60K_PCCP...1315792O}. Even then our formation efficiency is significantly higher than their calculated efficiency, which even includes an assumed solid-state dissociation efficiency of unity. We postulate that this is because of the difference in temperature and that the calculations only consider isolated events, while in the work described here many CO molecules and OH radicals are present at once, which could increase reaction probabilities.

A possible source of error in our determination of the formation efficiency are the experimental assumptions. Using Gaussian error propagation, we estimate the error in our formation efficiency of CO$_2$ per absorbed VUV photon in the reactive surface to be 60\%, and thus ranges from 7--27\%. Errors include, but are not limited to, assumptions made on the area which the infrared probes, the apparent band strength of $^{13}$C$^{18}$O$_2$, and the RAIRS correction factor. Although the propagation of all the individual errors results in a large uncertainty on our derived efficiency it is still at least four times smaller than the generally assumed value of unity for UV dissociation of water ice.

\begin{figure*}[ht!]
    \centering
    \includegraphics[width=\hsize]{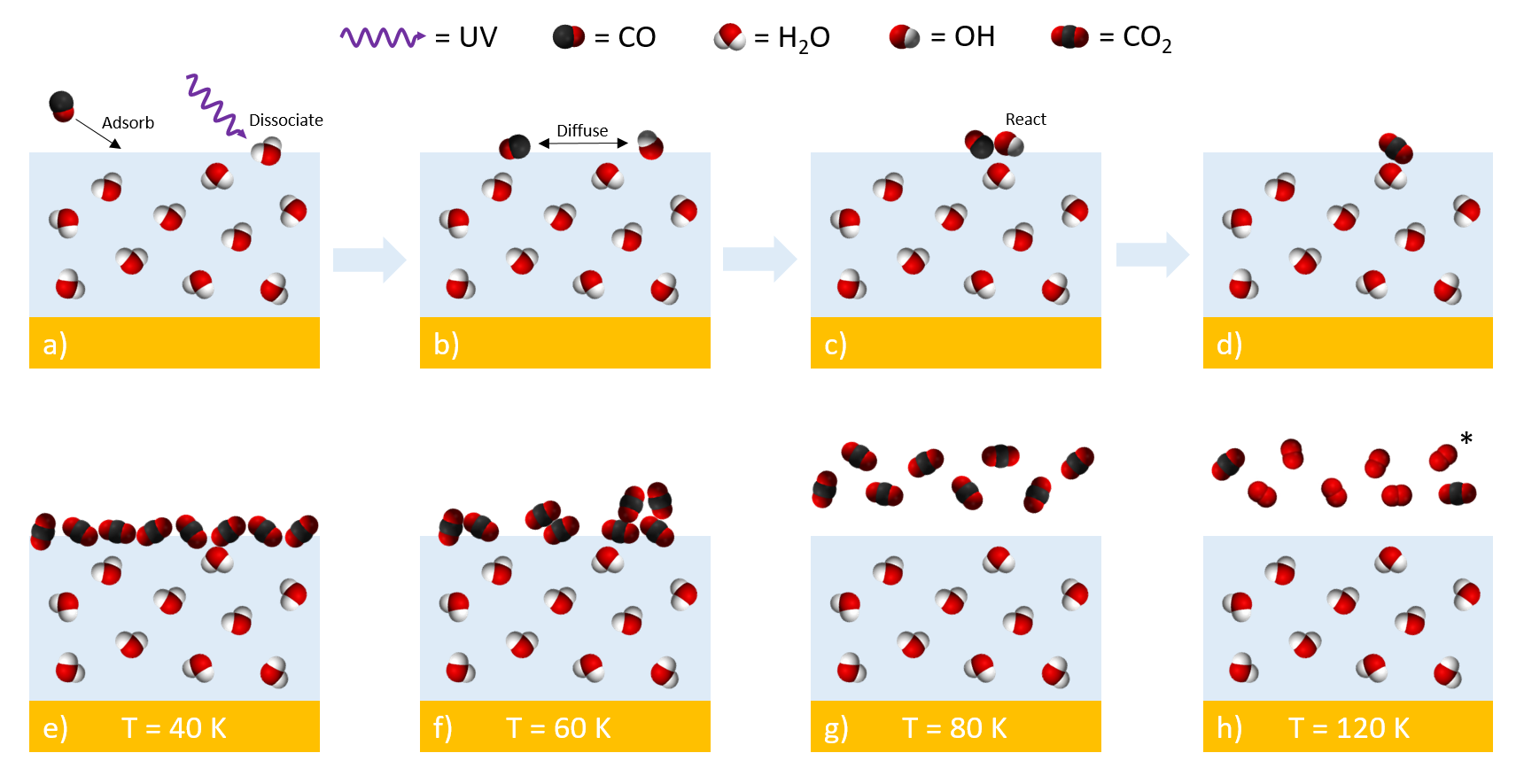}
    \caption{Simplistic representation of the processes that occur in our experiments. Panel a) through d) show the proposed Langmuir-Hinshelwood reaction that creates CO$_2$ in our experiments. Panels in the bottom row schematically show the structure or phase in which the formed products end up after formation, given a certain experimental temperature. $^{(*)}$ Molecular oxygen is not formed through the depicted reaction, for details see Sect. \ref{ssec:network} \& \ref{ssec:CO2vsO2}.} \label{fig:schematic}
\end{figure*}

\section{Astrophysical implications} \label{sec:astro_imp}
Figure~\ref{fig:schematic} schematically summarizes the above results and discussion. The formation of CO$_2$ through interaction between gas-phase CO and VUV induced OH radicals on water ice is visualized in the top row of this figure. Details on this process are discussed in Sect.~\ref{sec:disc}. In short, CO has a short, but nonzero, residence time on water ice even though its temperature is above the canonical sublimation temperature of CO. Concurrently, the water ice is irradiated with VUV photons, which results in the production of OH radicals. The diffusion of CO and OH radicals across the surface allows them to react with each other and form CO$_2$ in a Langmuir-Hinshelwood type reaction. Depending on the temperature of the water ice, the CO$_2$ is either mainly bound to the water ice, forms aggregates on top of the water ice, or is released into the gas phase, as is shown in the bottom row of Fig.~\ref{fig:schematic}. A full analysis of the experiments is given in Sect.~\ref{sec:results}, but briefly, in the lower end of the experimental temperature range, that is, 40--60~K, the formed CO$_2$ remains in the solid state. Specifically, at 40~K (Fig.~\ref{fig:schematic}e) the CO$_2$ is not mobile enough and the majority of the CO$_2$ stays bound to the ASW surface. However, in the experiment at 60~K (Fig.~\ref{fig:schematic}f), the CO$_2$ has significant mobility and starts diffusing across the surface, and forms CO$_2$ aggregates on top of the ASW. As shown in Fig.~\ref{fig:schematic}g, the formed CO$_2$ is released into the gas phase at 80~K, and at higher temperatures, the formation of O$_2$ starts competing with CO$_2$, which is shown in Fig.~\ref{fig:schematic}h. In the 40--90~K range, our experiments show that 7--27\% of the absorbed UV photons in the reactive surface, that is, top three monolayers of ASW, result in the conversion of gas-phase CO into CO$_2$. In the following section, we look at two astrophysical environments, that is, protoplanetary disks and molecular clouds, where this pathway could play an important role and may explain observational findings.

\subsection{CO conversion in protoplanetary disks} \label{ssec:ppds}
In planet forming disks, the gas mass, as derived through CO isotopologues, often comes out factors 10--100 lower than expected based on the dust content. This finding is based on physical-chemical that includes photodissociation and freeze out of CO and its isotopologues or thermo-chemical modeling \citep{2014_Miotello_CO_mass_A&A...572A..96M, 2016_Miotello_CO_mass_A&A...594A..85M, 2017_Miotello_Lupus_A&A...599A.113M, 2021_Calahan_TW_Hya_ApJ...908....8C}. One interpretation is that these disks have already lost a significant fraction of their total gas mass. Another is that some unknown process locks up gas-phase CO on grains. A correct interpretation is essential for models of planet formation that rely on the available gas-mass reservoir as well as on the gas-to-dust ratio. Other gas-mass tracers are problematic, as H$_2$ is undetectable and HD has only been observed in a few sources \citep[see e.g.,][]{2013_Bergin_HD_Natur.493..644B, 2016_McClure_HD_ApJ...831..167M, 2017_Trapman_HD_A&A...605A..69T, 2020_Kama_HD_A&A...634A..88K}, where, incidentally, the HD data support the notion of gas-phase CO being locked up. The CO into CO$_2$ conversion has been proposed as a possible pathway to convert gas-phase CO into a species that is much more difficult to detect.

Chemical models by \citet{2018_Bosman_CO_A&A...618A.182B} looked at several pathways through which CO could be converted into less volatile species to explain the low observed CO fluxes. These models are successful in this conversion on timescales shorter than average protoplanetary disk lifetimes, that is, $\sim3$~Myr. It should be noted that it was found in these models that gas-phase CO is in competition with atomic hydrogen for OH radicals on the surface \citep[see][for more details]{2018_Bosman_CO_A&A...618A.182B}. Furthermore, these results depend on the adopted binding energies, reaction rates, and formation of H$_2$. For example, the binding energy of CO in these models is kept constant at 855~K, no matter the environment. However, it has been found that the binding energy of CO on ASW can be as high as 1700~K \citep{2016_Jiao_bind_energy_ApJ...825...89H}. In a similar fashion, \citet{2021_Trapman_CO_conv_A&A...649A..95T} used physical-chemical models to investigate the low CO fluxes in the Lupus star-forming region. Disk regions with $T_{\mathrm{gas}}>35$~K were excluded in these models for gas-phase CO conversion, as verification models showed that CO conversion through grain-surface chemistry is negligible at these temperatures, but again this was tested with a CO binding energy of 855~K.

Including the correct binding energy for CO on ASW in these types of models is crucial. This increased binding energy allows gas-phase CO to compete with atomic hydrogen for OH radicals in a larger temperature range, and thus a larger region in protoplanetary disks where CO can be removed from the gas phase. In our experiments, the models by \citet{2018_Bosman_CO_A&A...618A.182B}, and \citet{2021_Trapman_CO_conv_A&A...649A..95T}, the amount of gas-phase CO that is converted into CO$_2$ depends on the availability of OH radicals. The authors assume in their models that once H$_2$O absorbs a VUV photon, dissociation is 100\% efficient. As discussed in Sect.~\ref{ssec:CO_into_CO2}, our experiments show an efficiency for OH production of 7--27\% per absorbed UV photon as opposed to the 100\% adopted in the models. Further modeling is required, which also includes the correct binding energy of CO on ASW, to fully assess the impact of our results on the model predictions about the consumption of gas-phase CO through this gas-grain interaction.

\subsection{CO$_2$ formation in edges of molecular clouds} \label{ssec:mol_clouds}
Observations of icy grains surrounding young stars suggest that large amounts of the observed solid-state CO$_2$ is in a water-rich environment \citep[see e.g.,][]{2015_Boogert_icy_universe_ARA&A..53..541B}. Here we explore if the conversion of gas-phase CO into solid-state CO$_2$ investigated in this work could (partially) explain the presence of solid-state CO$_2$ embedded in water ice \citep{2008_Pontoppidan_CO2_ApJ...678.1005P}. 

It is generally assumed that the ice that covers dust grains is composed of two layers: a polar and an apolar layer, where the apolar layer is formed on top of the polar layer . The polar ice layer is dominated by species with larger dipole moments, such as H$_2$O, and the apolar ice layer mainly contains species with smaller or no dipole moments, such as CO and N$_2$. Over the years solid-state CO$_2$ has been detected in a large number of sources. For example, \citet{1999_Gerakines_ISO_CO2_ApJ...522..357G} observed solid-state CO$_2$ with ISO in molecular clouds in a range of different physical environments. Their analysis shows that the majority of the observed solid-state CO$_2$ is found in a polar environment. Additionally, the \textit{Spitzer} ``Cores to Disks'' program similarly showed that in embedded young low-mass stars the majority of the observed solid-state CO$_2$ is mixed in a polar water-rich environment \citep{2008_Pontoppidan_CO2_ApJ...678.1005P}. Given these H$_2$O-dominated environments, it is possible that a substantial fraction of this CO$_2$ is formed through reactions between CO and OH radicals. This would require CO to freeze out during H$_2$O formation in order to be intimately mixed and subsequently to be converted into CO$_2$.

Infrared observations show that it is unlikely that CO is mixed in a water-rich environment. The observed CO absorption feature can be deconvolved into three components, namely 4.665~$\mu$m (2143.7~cm$^{-1}$), 4.673~$\mu$m (2139.7~cm$^{-1}$), and 4.681~$\mu$m (2136.5~cm$^{-1}$). The blue component (4.665~$\mu$m) is assigned to CO in an apolar environment, specifically, it is linked to mixtures of solid-state CO and CO$_2$ \citep{2002_Boogert_CO_ApJ...568..761B, 2006_Broekhuizen_CO-CO2_A&A...451..723V} or crystalline CO \citep{2003_Pontoppidan_CO_A&A...408..981P}. The middle component (4.673~$\mu$m) is generally attributed to pure CO \citep[see e.g.,][]{2002_Boogert_CO_ApJ...568..761B}. The red component (4.681~$\mu$m) has a broader ``footprint'' compared to the other two and is linked to CO in a polar environment. This polar environment could be H$_2$O and would set the scene for solid-state formation of CO$_2$ in a polar environment. However, \citet{1988_Sandford_CO_ApJ...329..498S} showed that if CO would reside in H$_2$O, one can expect a feature at 4.647~$\mu$m (2151.9~cm$^{-1}$) due to the dangling OH bond, which has not been seen in interstellar spectra. It is thus unlikely that this CO is mixed with H$_2$O. Mixtures of CO with CH$_3$OH, however, do reproduce the red component in both peak position and width \citep{2011_Cuppen_CO_CH3OH_MNRAS.417.2809C}. A mixture of CO and CH$_3$OH is also more likely, as CH$_3$OH is formed through hydrogenation of CO \citep{1994_Hiraoka_hydro_CPL...229..408H, 2002_Hiraoka_hydro_ApJ...577..265H, 2002_Watanabe_hydro_ApJ...571L.173W, 2004_Hidaka_H2CO_ApJ...614.1124H, 2004_Watanabe_hydro_ApJ...616..638W, 2009_Fuchs_hydro_A&A...505..629F, 2022_Santos_CH3O_H2CO_ApJ...931L..33S}.

In order to explain the solid-state CO$_2$ embedded in water-rich environments without invoking the need for CO embedded in H$_2$O ice, we look at the initial build-up of water ice on dust grains. Water can be formed through both gas-phase and solid-state pathways. However, the gas-phase ion-molecule chemistry produces only a fraction of the total observed water abundance, and thus water is mainly formed through the addition of hydrogen to atomic oxygen on the surface of dust grains \citep[see reviews by][and references therein]{2014_vanDishoeck_H2O_prpl.conf..835V, 2015_Linnartz_review_arXiv150702729L}. This formation process of solid-state H$_2$O takes place in the edges of molecular clouds at intermediate extinction ($A_V$). At this extinction, CO is already present, but still resides in the gas phase. For example, in the Taurus molecular cloud water ice is detected at a threshold extinction, that is, the extinction at which a species is detected in the solid state, of $3.2\pm0.1$ $A_V$, while for CO the threshold extinction was determined to be $6.7\pm1.6$~$A_V$ \citep{2001_WHittet_thresAv_H2O_ApJ...547..872W, 2010_Whittet_CO_function_Av_ApJ...720..259W}. Physical-chemical models of molecular clouds show that the dust-grain temperature at the edge of a cloud equals 31 K, with an external VUV field strength of 100~$G_\mathrm{0}$, where $G_\mathrm{0}$ is a scaling factor in multiples of the average local interstellar radiation field \citep{1968_Habing_G0_BAN....19..421H}, which indeed is sufficient to keep CO in the gas phase \citep{2009_Hollenbach_MC_models_ApJ...690.1497H}. For CO$_2$ to be mixed with H$_2$O it has to form simultaneously, and since CO is in the gas-phase during H$_2$O formation, CO$_2$ can be formed through the process described in this work. The OH radicals required for the conversion of CO into CO$_2$ are readily available in this region. They are the intermediate product to H$_2$O formation, and the external UV field is still sufficient at this extinction to photodissociate already existing H$_2$O molecules.

The solid-state CO$_2$ that is observed to be embedded in water-rich environment in the ``Cores to Disks'' program, has a relative average abundance w.r.t. H$_2$O of $\sim0.2$. Assuming that roughly equal amounts of atomic oxygen go into CO and H$_2$O, this would imply that approximately 20\% of the gas-phase CO would have to be converted into solid-state CO$_2$ to explain observed abundances. Inclusion of this pathway in physical-chemical models of molecular clouds is required to test how efficient this process is in low-density regions at interstellar timescales and explain the observed solid-state CO$_2$ abundances.

\section{Conclusions} \label{sec:conc}
The work presented is principally different from many other solid-state astrochemical experiments presented in the past, as it implicitly takes into account gas-grain interactions. A number of systematic measurements have been performed in the temperature range of 40--120~K. Control experiments are performed to narrow down the possible interpretations of the results. Our findings are summarized as follows:
\begin{enumerate}
    \item The interaction between gas-phase CO and vacuum-UV irradiated water ice produces CO$_2$ up to 120~K. Solid-state CO$_2$ is observed in the temperature range 40--60 K. At 70~K or above, the formed CO$_2$ is released into the gas phase. Additionally, above 90 K the formation of O$_2$ is observed at the cost of CO$_2$ production.
    \item The residence time of CO on water ice is significant, even though it is above the canonical sublimation temperature of CO. In this short, but nonzero, residence time, a CO molecule is able to diffuse up to 7.2 $\times$ 10$^{10}$ different binding sites before desorption occurs. This significant diffusion allows CO to find an OH radical, created by VUV dissociation of H$_2$O, and form CO$_2$ in a Langmuir-Hinshelwood type reaction.
    \item Given that gas-phase CO can only interact with the surface of the water ice, this includes pores exposed to the vacuum, we derived a conversion efficiency of 7--27\% per absorbed photon in the reactive surface (i.e., top three monolayers). The limiting factor in this conversion rate is the production of OH radicals.
    \item The VUV dissociation efficiency of solid-state H$_2$O is likely the limiting factor in the above conversion efficiency from gas-phase CO into CO$_2$.
    \item Understanding this process is important for astrophysical regions, such as planet-forming disks and molecular clouds. In clouds, this process can explain the presence of solid-state CO$_2$ embedded in water-rich ices. In disks, it has been invoked to explain the lack of gas-phase CO. Our results suggest that the process might be more complex then those incorporated in the physical-chemical models. Further theoretical investigation is required to investigate the conversion of gas-phase CO into CO$_2$ to its full extent.
\end{enumerate}

This work demonstrates the wide temperature efficacy of this gas-grain interaction process. Future work, should focus on further experimental and theoretical exploration of the molecular dynamics which include the effects of high fluxes and neighbouring OH radicals on the reforming of H$_2$O after photodissociation. Additionally, these results should be included quantitatively in models of planet-forming disks and molecular clouds. With this work we show that gas and grain chemistry cannot be considered as fully separate, but that, under the right conditions, interaction of the gas with the icy surface results in observable effects. It is interesting to realize, that similar processes might also at play for other gas-phase molecules that interact with icy grains.

\begin{acknowledgements}
The authors thank the referee for the constructive feedback. We also thank E. van Dishoeck, C. Eistrup, and C. Walsh for their participation in insightful discussions. This research was funded through the Dutch Astrochemistry II program of the Netherlands Organization for Scientific Research (648.000.025) and NOVA, the Netherlands Research School for Astronomy. JTvS acknowledges recent financial support through the Virginia Initiative on Cosmic Origins (VICO) postdoctoral fellowship program and NL through an SNSF Ambizione grant (\#193453).
\end{acknowledgements}

\bibliography{JTvS_CO_OH}
\bibliographystyle{aa}

\begin{appendix}

\section{VUV spectrum} \label{app:uv_spec}
\begin{figure}[t!]
    \includegraphics[width=\hsize]{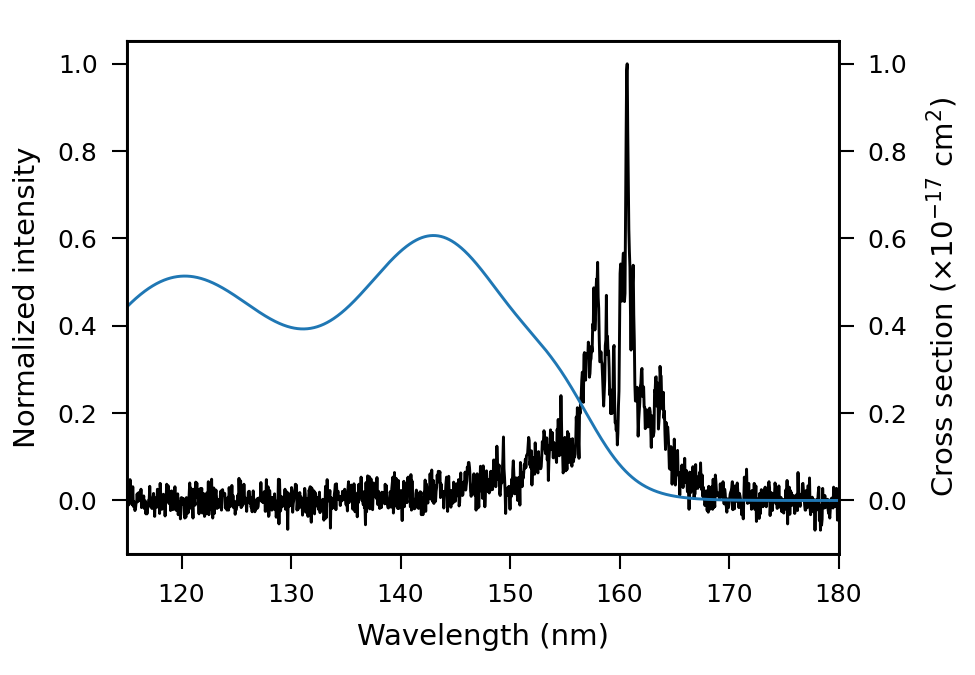}
    \caption{Black: In-situ measured VUV spectrum of the microwave-discharge hydrogen-flow lamp used in combination with a MgF$_2$ window that absorbs Lyman-$\alpha$. Blue: solid-state H$_2$O VUV absorption cross section as function of wavelength taken from \citet{2014_Cruz-Diaz_VUV_cross_A&A...562A.119C}.} \label{fig:uv_spec}
\end{figure}

The spectral energy distribution of the MDHL used in this study is measured in situ with a VUV spectrometer (McPherson Model 234/302), which is mounted opposite to the MDHL on the other side of the main chamber. In this work a MgF$_2$ window was used that does not transmit Lyman-$\alpha$ photons, but does transmit the molecular hydrogen emission lines and continuum between 140--170 nm. The VUV absorption cross section as a function of wavelength is taken from literature \citep{2014_Cruz-Diaz_VUV_cross_A&A...562A.119C}. It is the summation of three Gaussians with the parameters taken from their Table~2. Both the VUV spectrum with which the ASW is irradiated and the wavelength dependent H$_2$O VUV absorption cross section are shown in Fig. \ref{fig:uv_spec}.

\section{Additional experiments} \label{app:add_exp}
\begin{figure*}[t!]
    \includegraphics[width=\hsize]{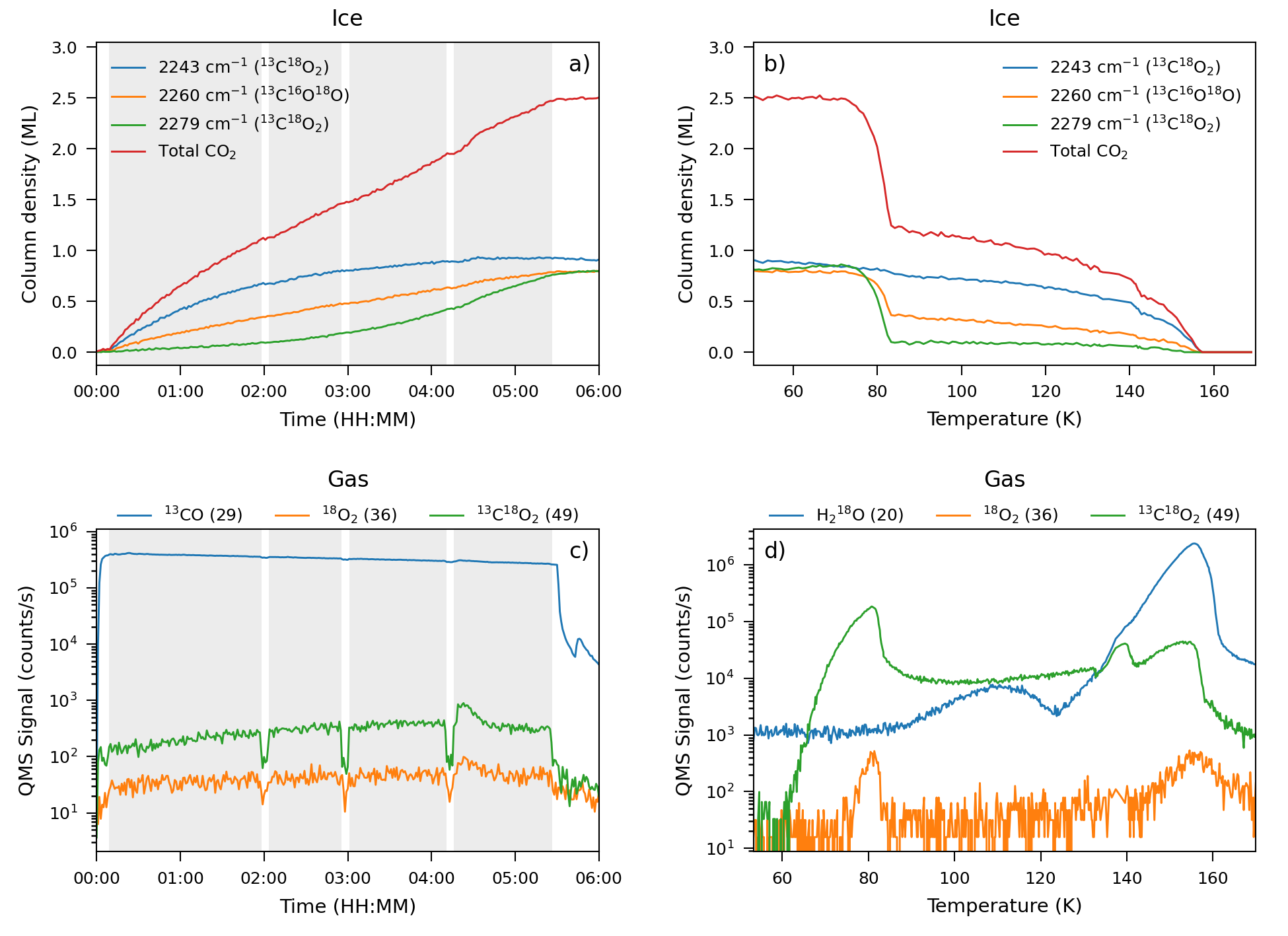}
    \caption{Results of the experiment of ASW at a temperature of 50 K. In the top row the deconvolved infrared components, which trace the solid state, are given, based on RAIRS during VUV irradiation (panel a) and temperature programmed desorption (TPD) in panel b). In the bottom row the data from the quadrupole mass spectrometer (QMS) is presented, which trace the gas phase. The left column shows data acquired during VUV irradiation (shaded areas indicate when the VUV shutter is open) and on the right during TPD. In panel c) $^{13}$CO is shown, as measuring the main isotopologue ($^{13}$C$^{18}$O) would saturate the QMS. The $^{16}$O isotope is present in the CO sample at a level of 5\%. The TPD QMS data in panel d) between 130--140 K is unreliable due to a nonlinear temperature artifact during heating of the sample.}
    \label{fig:exp_50K}
\end{figure*}

The main experiments at 50, 60, 80, 90, 100, and 120 K and control experiments with $^{13}$CO instead of $^{13}$C$^{18}$O, no VUV irradiation, and no gas-phase CO in the chamber during VUV irradiation are presented in this appendix. In the majority of the experiments there is an artifact in the QMS data during TPD, as can be seen in for example Fig. \ref{fig:exp_50K}d. This is the result of a nonlinear temperature increase in the sample, which is likely caused by a change in thermal conductivity or contact between the sample and heating strip. The LakeShore PID controller compensates for this, but results in a brief period of undershoot followed by overshoot of the sample temperature. This occurs approximately in the temperature range 120--150 K. In the caption of each figure the exact temperature range is given in which the QMS signals are not accurately representing what the TPD should look like if the temperature increase were to be fully linear.

\begin{figure*}[t!]
    \includegraphics[width=\hsize]{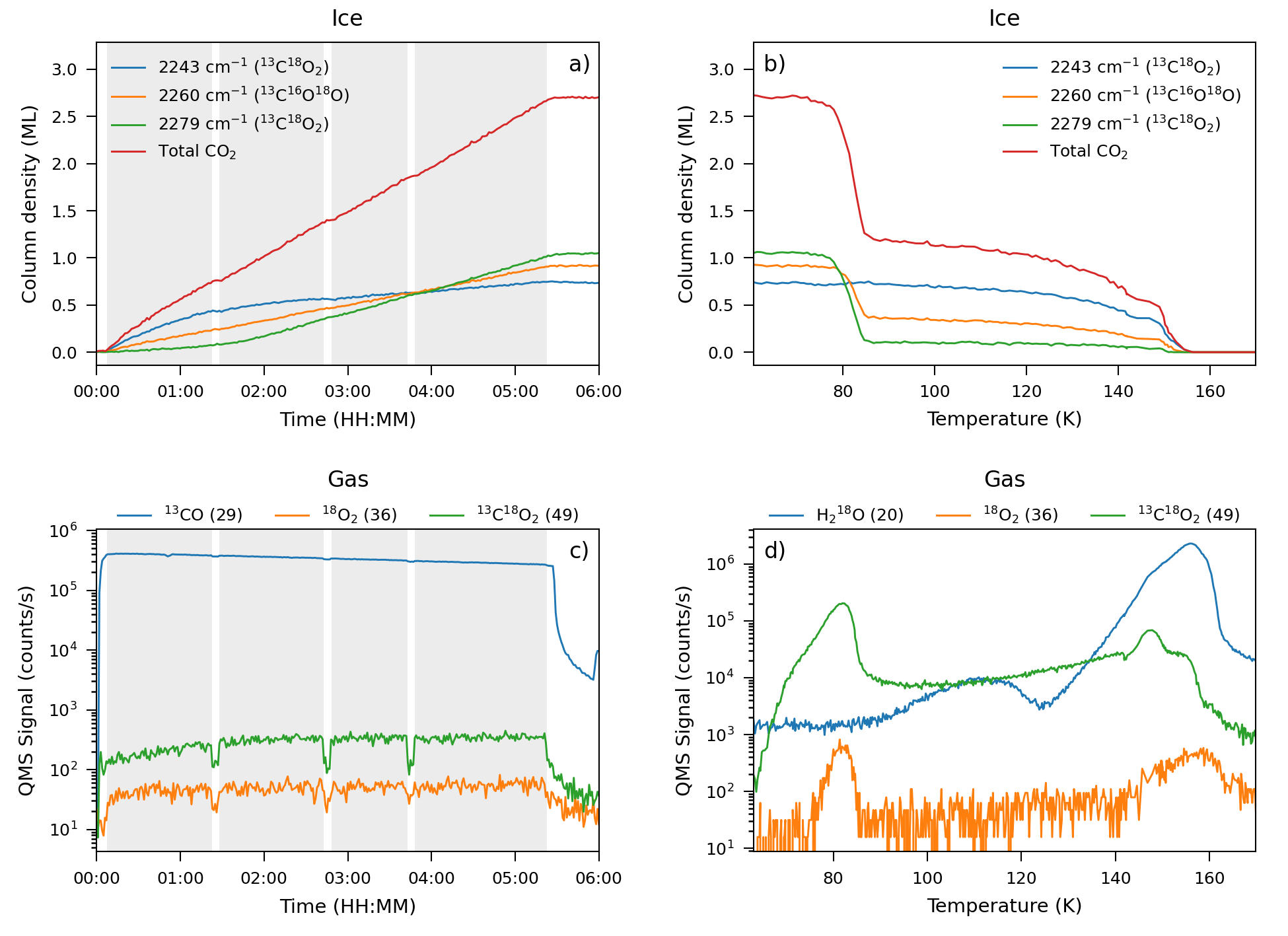}
    \caption{Results of the experiment of ASW at a temperature of 60 K. In the top row the deconvolved infrared components, which trace the solid state, are given, based on RAIRS during VUV irradiation (panel a) and temperature programmed desorption (TPD) in panel b). In the bottom row the data from the quadrupole mass spectrometer (QMS) is presented, which trace the gas phase. The left column shows data acquired during VUV irradiation (shaded areas indicate when the VUV shutter is open) and on the right during TPD. In panel c) $^{13}$CO is shown, as measuring the main isotopologue ($^{13}$C$^{18}$O) would saturate the QMS. The $^{16}$O isotope is present in the CO sample at a level of 5\%. The TPD QMS data in panel d) between 140--150 K is unreliable due to a nonlinear temperature artifact during heating of the sample.}
    \label{fig:exp_60K}
\end{figure*}

\begin{figure*}[t!]
    \includegraphics[width=\hsize]{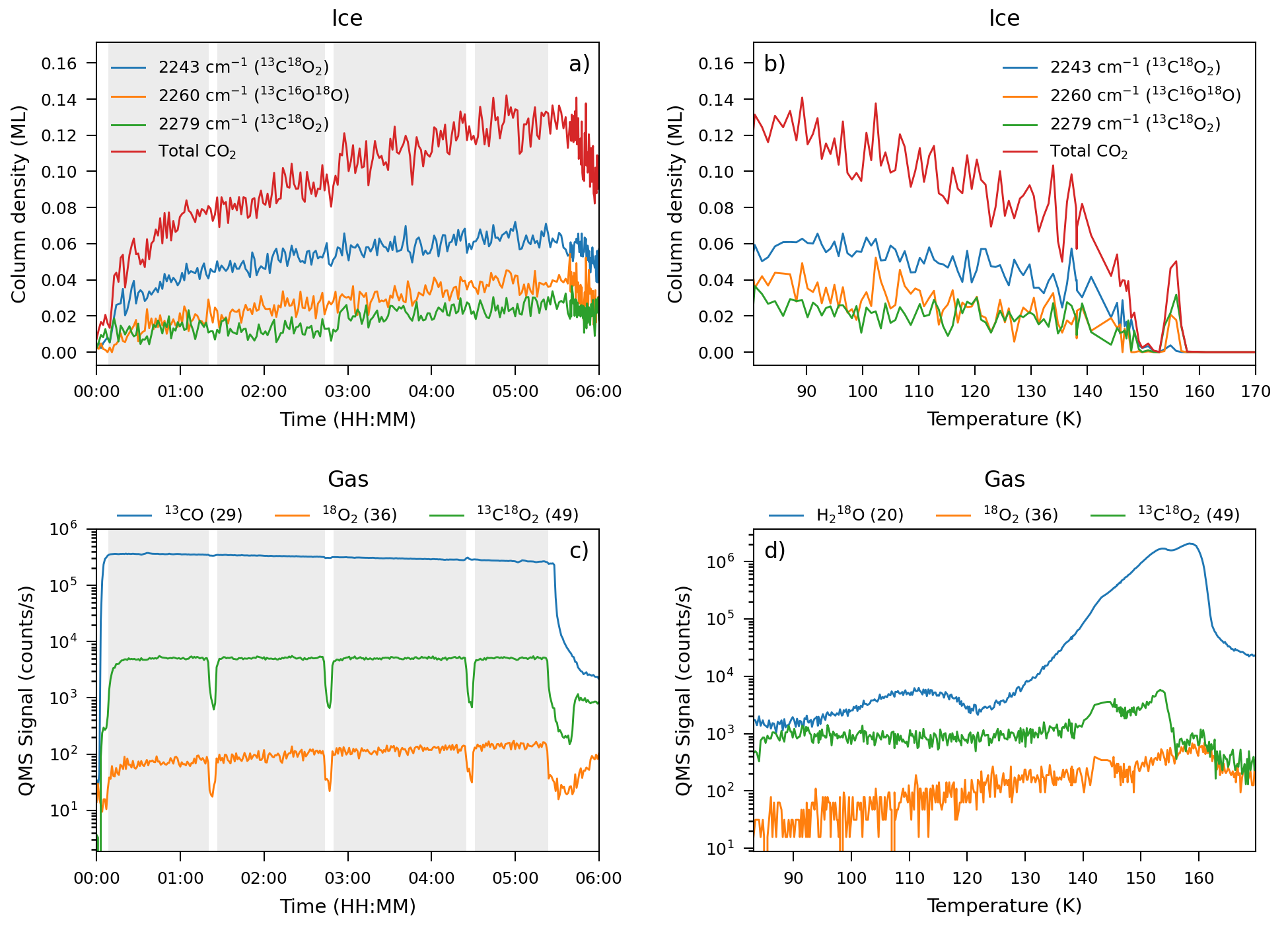}
    \caption{Results of the experiment of ASW at a temperature of 80 K. In the top row the deconvolved infrared components, which trace the solid state, are given, based on RAIRS during VUV irradiation (panel a) and temperature programmed desorption (TPD) in panel b). In the bottom row the data from the quadrupole mass spectrometer (QMS) is presented, which trace the gas phase. The left column shows data acquired during VUV irradiation (shaded areas indicate when the VUV shutter is open) and on the right during TPD. In panel c) $^{13}$CO is shown, as measuring the main isotopologue ($^{13}$C$^{18}$O) would saturate the QMS. The $^{16}$O isotope is present in the CO sample at a level of 5\%. The TPD QMS data in panel d) between 140--150 K is unreliable due to a nonlinear temperature artifact during heating of the sample.}
    \label{fig:exp_80K}
\end{figure*}

\begin{figure*}[t!]
    \includegraphics[width=\hsize]{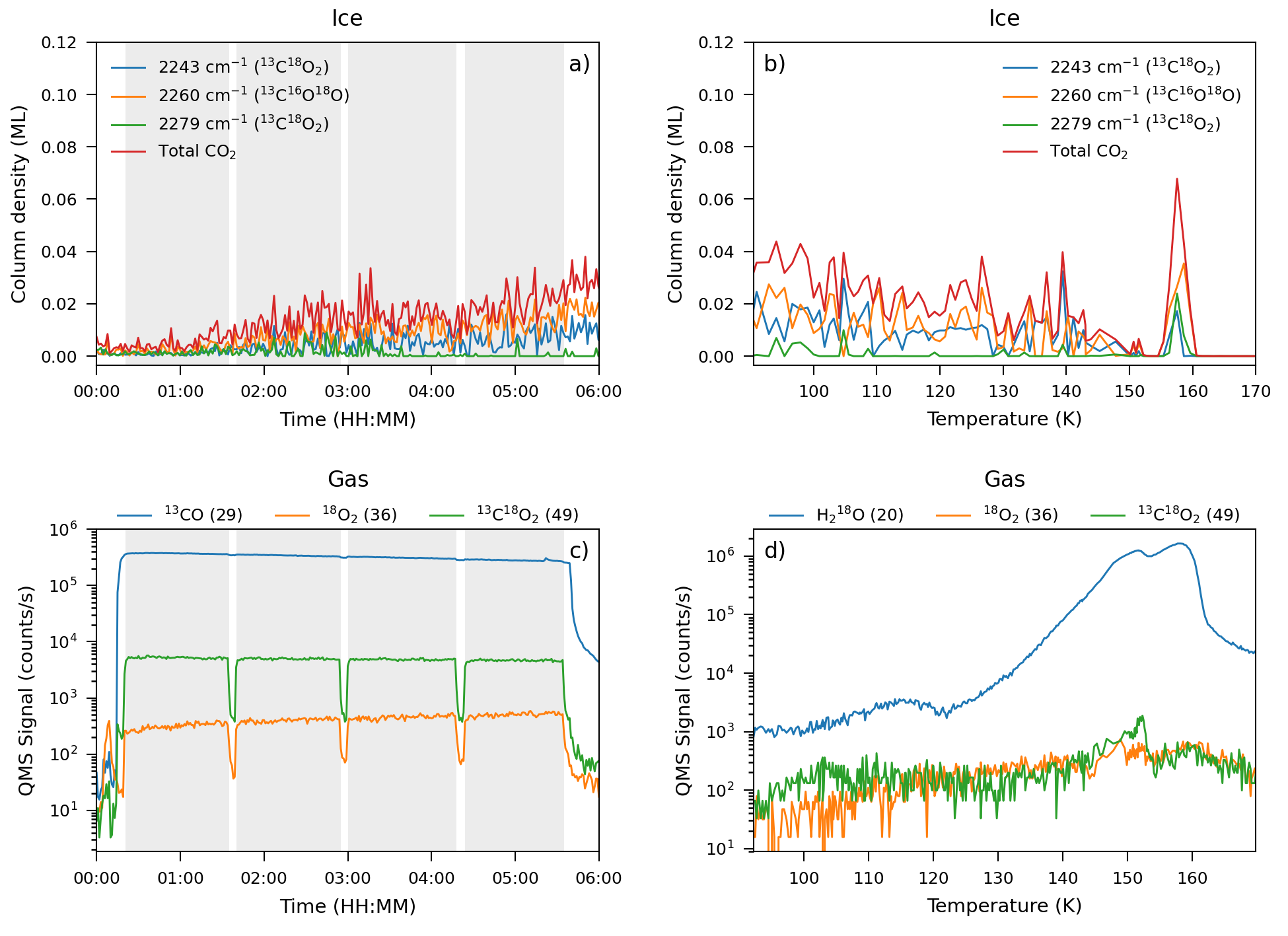}
    \caption{Results of the experiment of ASW at a temperature of 90 K. In the top row the deconvolved infrared components, which trace the solid state, are given, based on RAIRS during VUV irradiation (panel a) and temperature programmed desorption (TPD) in panel b). In the bottom row the data from the quadrupole mass spectrometer (QMS) is presented, which trace the gas phase. The left column shows data acquired during VUV irradiation (shaded areas indicate when the VUV shutter is open) and on the right during TPD. In panel c) $^{13}$CO is shown, as measuring the main isotopologue ($^{13}$C$^{18}$O) would saturate the QMS. The $^{16}$O isotope is present in the CO sample at a level of 5\%.}
    \label{fig:exp_90K}
\end{figure*}

\begin{figure*}[t!]
    \includegraphics[width=\hsize]{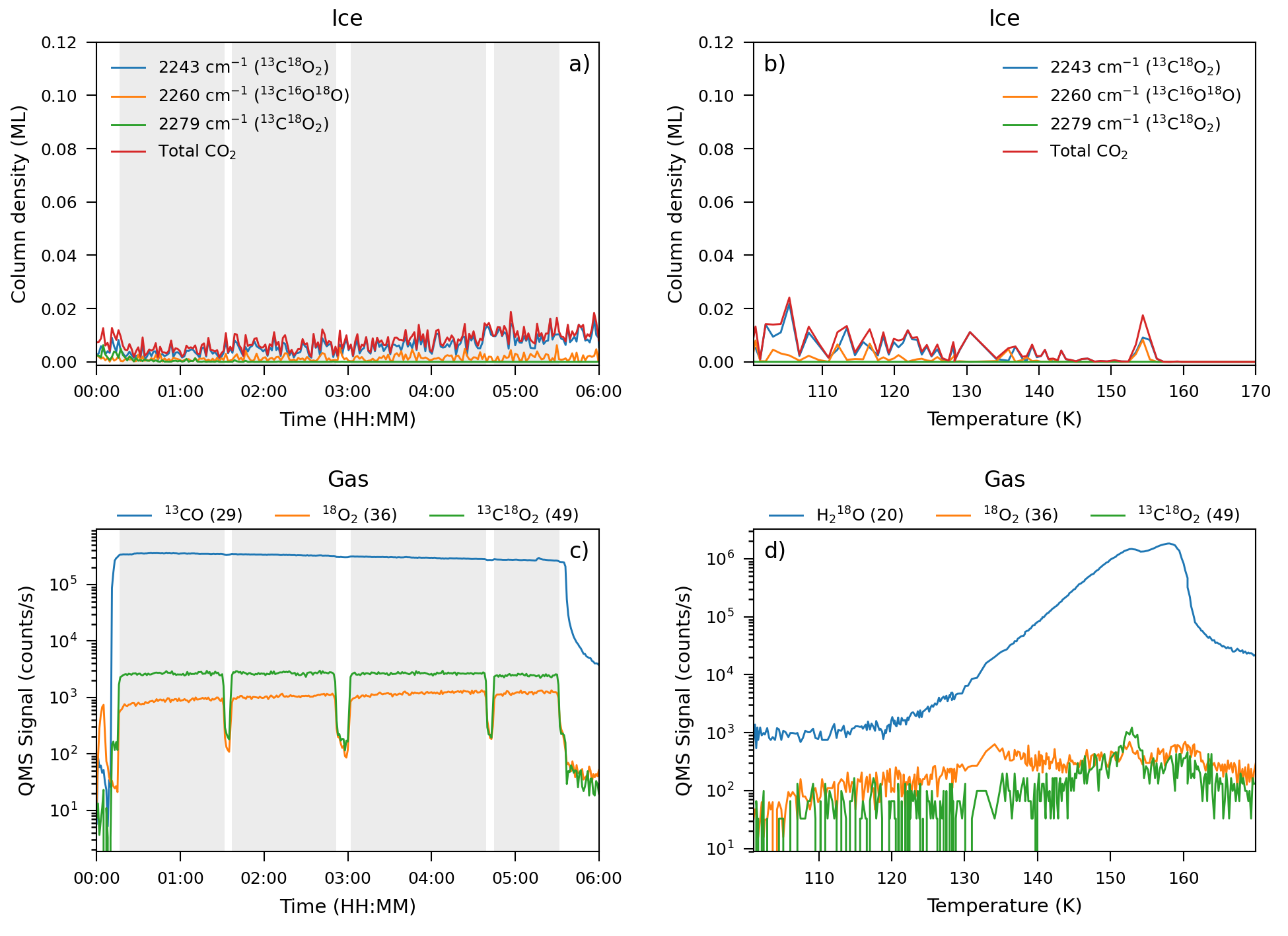}
    \caption{Results of the experiment of ASW at a temperature of 100 K. In the top row the deconvolved infrared components, which trace the solid state, are given, based on RAIRS during VUV irradiation (panel a) and temperature programmed desorption (TPD) in panel b). In the bottom row the data from the quadrupole mass spectrometer (QMS) is presented, which trace the gas phase. The left column shows data acquired during VUV irradiation (shaded areas indicate when the VUV shutter is open) and on the right during TPD. In panel c) $^{13}$CO is shown, as measuring the main isotopologue ($^{13}$C$^{18}$O) would saturate the QMS. The $^{16}$O isotope is present in the CO sample at a level of 5\%. The TPD QMS data in panel d) between 130--140 K is unreliable due to a nonlinear temperature artifact during heating of the sample.}
    \label{fig:exp_100K}
\end{figure*}

\begin{figure*}[t!]
    \includegraphics[width=\hsize]{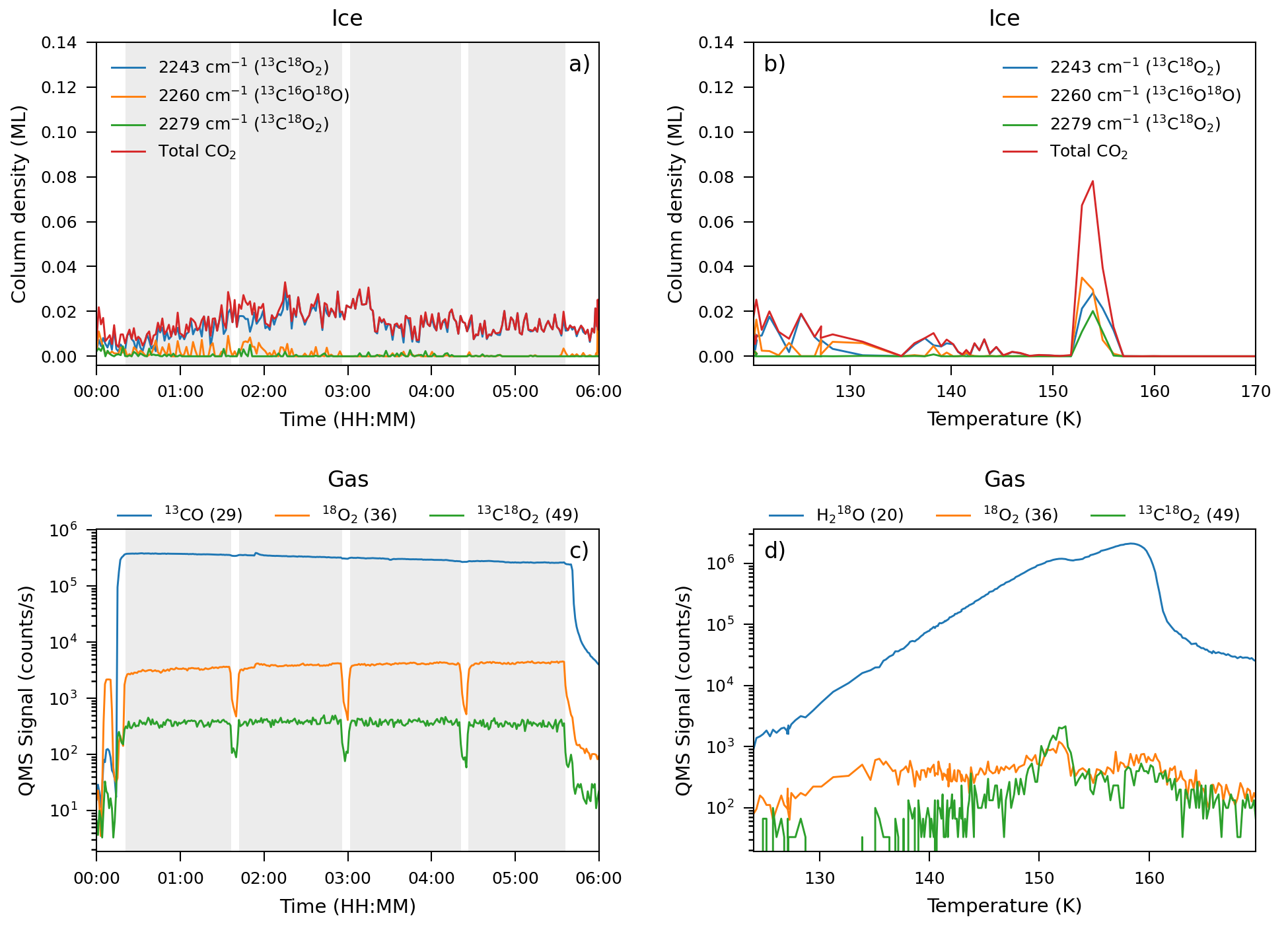}
    \caption{Results of the experiment of ASW at a temperature of 120 K. In the top row the deconvolved infrared components, which trace the solid state, are given, based on RAIRS during VUV irradiation (panel a) and temperature programmed desorption (TPD) in panel b). In the bottom row the data from the quadrupole mass spectrometer (QMS) is presented, which trace the gas phase. The left column shows data acquired during VUV irradiation (shaded areas indicate when the VUV shutter is open) and on the right during TPD. In panel c) $^{13}$CO is shown, as measuring the main isotopologue ($^{13}$C$^{18}$O) would saturate the QMS. The $^{16}$O isotope is present in the CO sample at a level of 5\%. The TPD QMS data in panel d) between 130--140 K is unreliable due to a nonlinear temperature artifact during heating of the sample.}
    \label{fig:exp_120K}
\end{figure*}

\begin{figure*}[t!]
    \includegraphics[width=\hsize]{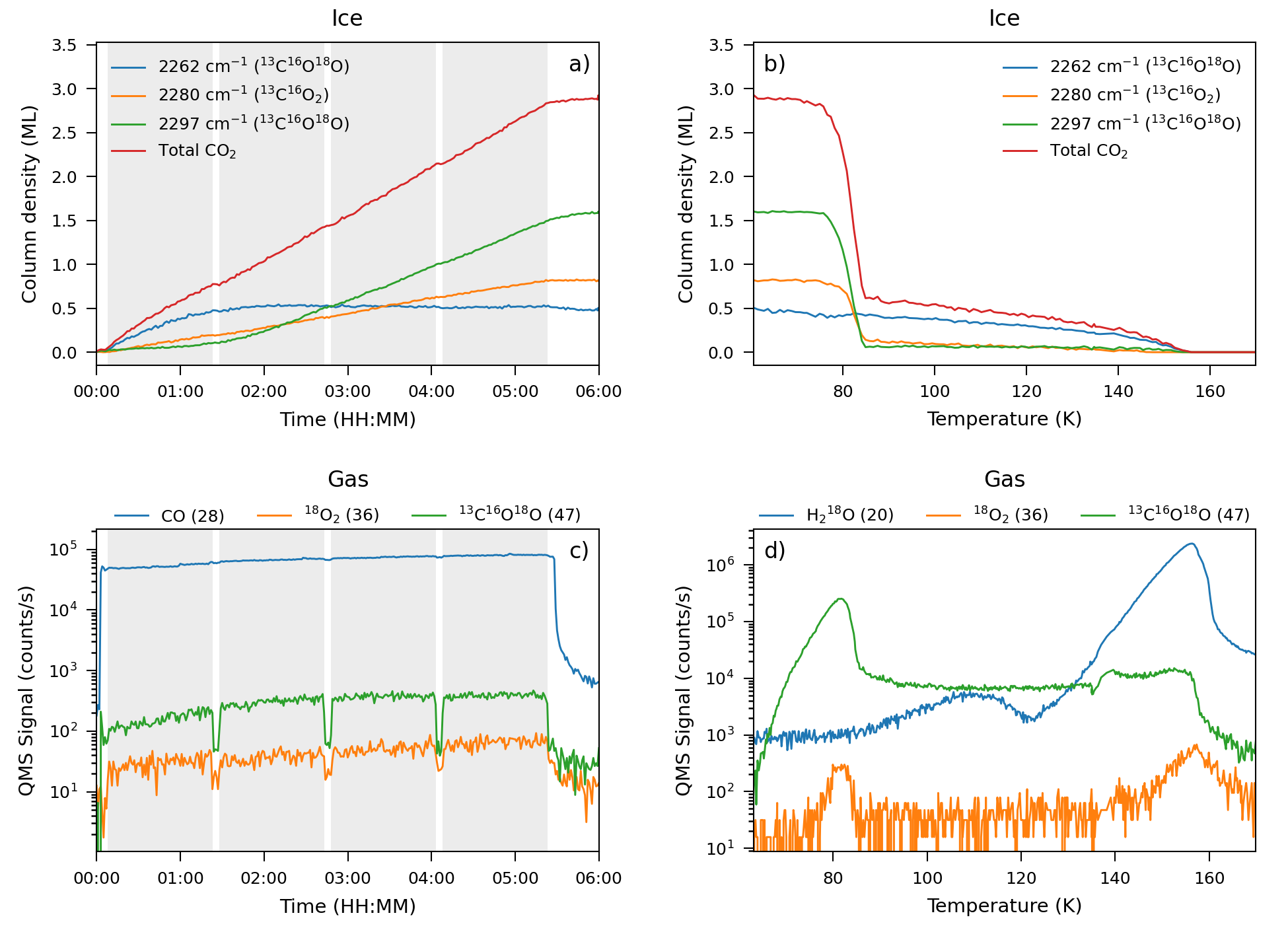}
    \caption{Results of the experiment of ASW at a temperature of 60 K with the isotopologue $^{13}$C$^{16}$O instead of $^{13}$C$^{18}$O. In the top row the deconvolved infrared components, which trace the solid state, are given, based on RAIRS during VUV irradiation (panel a) and temperature programmed desorption (TPD) in panel b). In the bottom row the data from the quadrupole mass spectrometer (QMS) is presented, which trace the gas phase. The left column shows data acquired during VUV irradiation (shaded areas indicate when the VUV shutter is open) and on the right during TPD. In panel c) $^{12}$CO is shown, as measuring the main isotopologue ($^{13}$CO) would saturate the QMS. The $^{12}$C isotope is present in the CO sample at a level of 1\%. The TPD QMS data in panel d) between 130--140 K is unreliable due to a nonlinear temperature artifact during heating of the sample.}
    \label{fig:exp_isotope}
\end{figure*}

\begin{figure*}[t!]
    \includegraphics[width=\hsize]{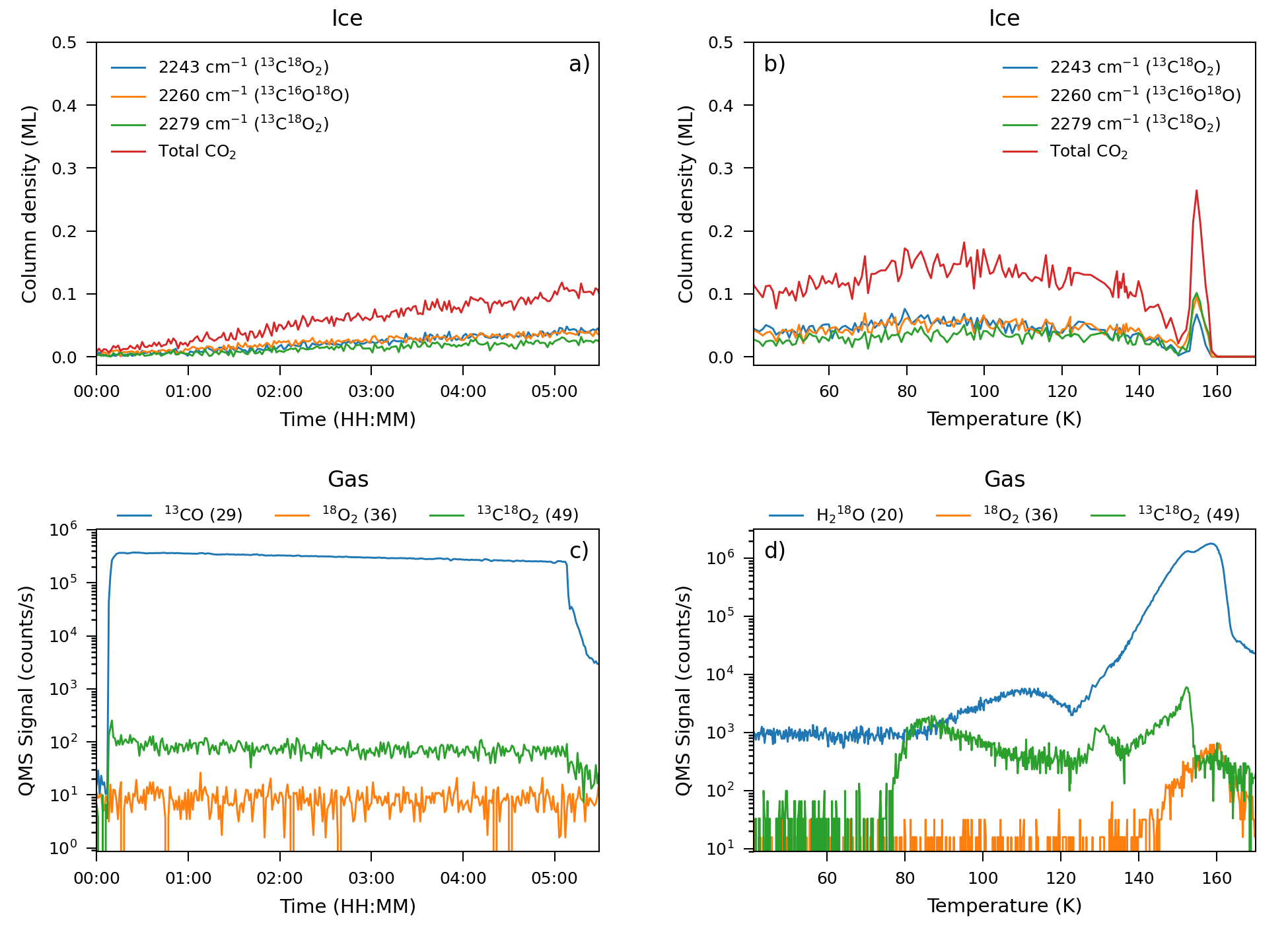}
    \caption{Results of the experiment of ASW at a temperature of 40 K without VUV irradiation. In the top row the deconvolved infrared components, which trace the solid state, are given, based on RAIRS during VUV irradiation (panel a) and temperature programmed desorption (TPD) in panel b). In the bottom row the data from the quadrupole mass spectrometer (QMS) is presented, which trace the gas phase. The left column shows data acquired during VUV irradiation (shaded areas indicate when the VUV shutter is open) and on the right during TPD. In panel c) $^{13}$CO is shown, as measuring the main isotopologue ($^{13}$C$^{18}$O) would saturate the QMS. The $^{16}$O isotope is present in the CO sample at a level of 5\%.}
    \label{fig:exp_no_uv}
\end{figure*}

\begin{figure*}[t!]
    \includegraphics[width=\hsize]{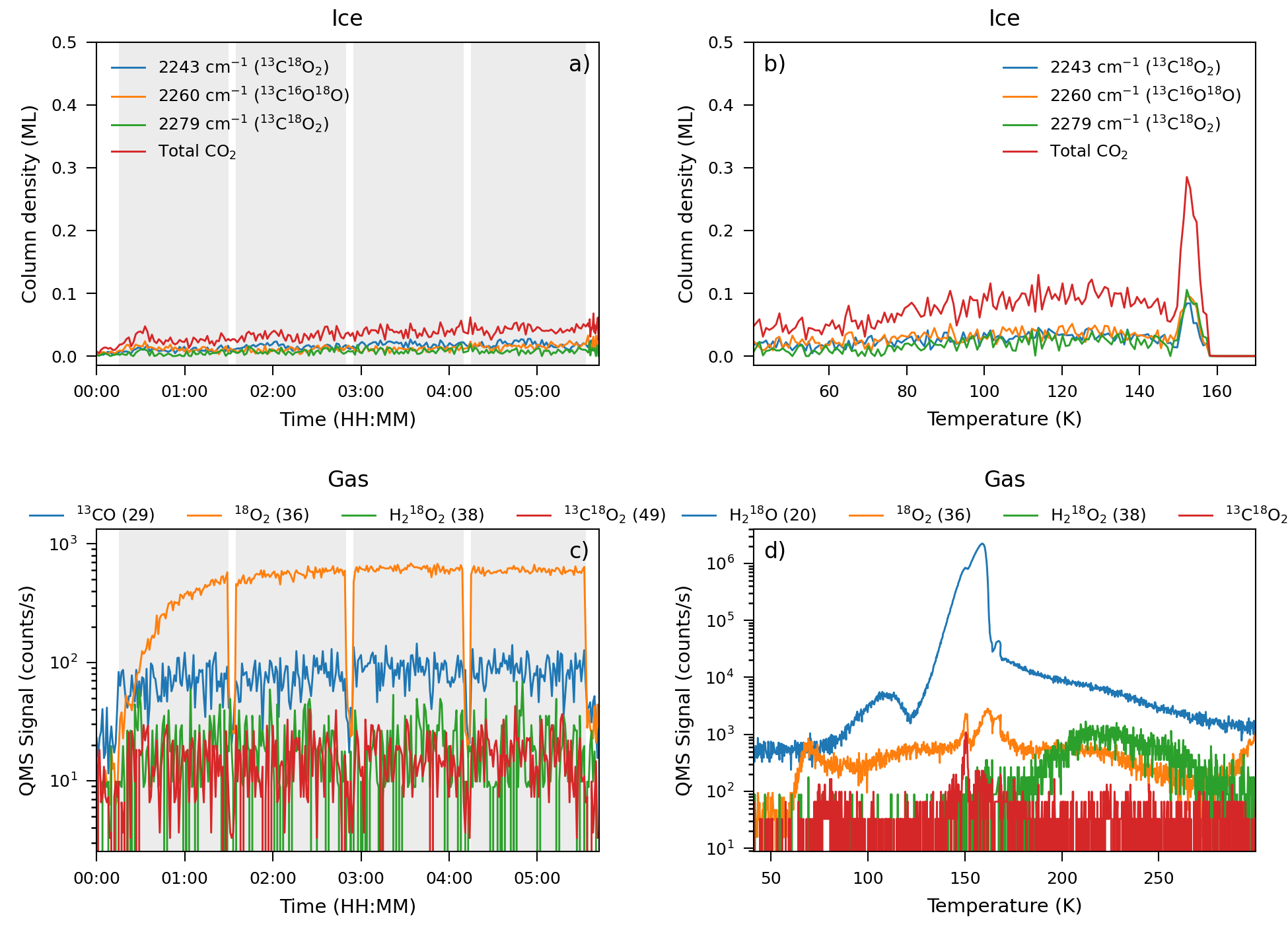}
    \caption{Results of the experiment of ASW at a temperature of 40 K without gas-phase CO. In the top row the deconvolved infrared components, which trace the solid state, are given, based on RAIRS during VUV irradiation (panel a) and temperature programmed desorption (TPD) in panel b). In the bottom row the data from the quadrupole mass spectrometer (QMS) is presented, which trace the gas phase. The left column shows data acquired during VUV irradiation (shaded areas indicate when the VUV shutter is open) and on the right during TPD. TPD is now shown up until 300 K to show the desorption of H$_2$O$_2$. In panel c) $^{13}$CO is shown, to verify that indeed no significant amount of CO is in the gas phase during VUV irradiation.}
    \label{fig:exp_water_only}
\end{figure*}

\end{appendix}
\end{document}